\documentclass[reprint,aps,pre]{revtex4-1}

\usepackage{amsmath,amsfonts,amscd,amssymb,graphicx}

\newcommand{\bb}{\mathbf{b}}

\newcommand{\bv}{\mathbf{v}}

\newcommand{\bx}{\mathbf{x}}

\newcommand{\bX}{\mathbf{X}}

\newcommand{\bfeta}{\boldsymbol{\eta}}
\newcommand{\bmu}{\boldsymbol{\mu}}

\newcommand{\bphi}{\boldsymbol{\phi}}
\newcommand{\bPhi}{\boldsymbol{\Phi}}

\DeclareMathAlphabet{\mathpzc}{OT1}{pzc}{m}{it}

\newcommand{\bG}{\mathbf{G}}

\DeclareMathOperator*{\argmin}{arg\,min}

\usepackage{dcolumn}% Align table columns on decimal point
\usepackage{overpic}
\usepackage{color}
\usepackage{adjustbox}
\usepackage[caption=false]{subfig}
\usepackage{rotating}
\usepackage{float}
\makeatletter
\let\newfloat\newfloat@ltx
\makeatother
\usepackage{algorithm}
\usepackage{algpseudocode}
\algnewcommand\algorithmicinput{\textbf{Input:}}
\algnewcommand\Input{\item[\algorithmicinput]}

\begin{document}

\title{Pilot-Wave Dynamics: Using Dynamic Mode Decomposition to characterize\\ Bifurcations, Routes to Chaos and Emergent Statistics}

\author{J. Nathan Kutz$^*$, Andr\'e Nachbin$^{**}$, Peter J. Baddoo$^\dag$ and John W. M. Bush$^\dag$}
 \affiliation{$^*$Department of Applied Mathematics and Electrical and Computer Engineering, University of Washington, Seattle, WA} 
 \affiliation{$^{**}$ Instituto de Matem\'atica Pura e Aplicada, Rio de Janeiro, Brazil}
 \affiliation{$^\dag$ Department of Mathematics, Massachusetts Institute of Technology, Cambridge, MA}

\begin{abstract}
We develop a data-driven characterization of the pilot-wave hydrodynamic system in which a bouncing droplet self-propels along the surface of a vibrating bath. We consider drop motion in a confined one-dimensional geometry, and apply the {\em Dynamic mode decomposition} (DMD) in order to characterize the evolution of the wave field as the bath's vibrational acceleration is increased progressively.  DMD provides a regression framework for adaptively learning a best-fit linear dynamics model over snapshots of spatio-temporal data.  The DMD characterization of the wave field yields a fresh perspective on the bouncing-droplet problem that forges valuable new links with the mathematical machinery of quantum mechanics.  Moreover, it provides a low-rank characterization of the bifurcation structure of the pilot wave physics.  Specifically, the analysis shows that as the vibrational acceleration is increased, the pilot-wave field undergoes a series of Hopf bifurcations that ultimately lead to a chaotic wave field.  
%The period-doubling route to chaos is a canonical observation of damped-driven systems.
The established relation between the mean pilot-wave field and the droplet statistics allows us to characterize the evolution of the emergent statistics with increased vibrational forcing from the evolution of the pilot-wave field.
We thus develop a numerical framework with the same basic structure as quantum mechanics, specifically
a wave theory that predicts particle statistics.
%We thus develop a theory, albeit numerically based, with the same structure as quantum mechanics, a wave theory that predicts particle statistics.
\end{abstract}
\maketitle

%%%%%%%%%%%%%%%%%%%%%%%%%%%%%%%%%%%%%%%%%%%%%%%%%%%%%%%%%%%%%%%%%%
\section{INTRODUCTION} \label{intro}

When studying quantum mechanics for the first time, one is struck by the fact that it is a wave theory that predicts the statistical behavior of particles. Specifically, the statistical behavior of particles may be deduced directly from the wavefunction, as it evolves according to the linear Schr\"odinger equation, without knowledge of the underlying particle dynamics. Attempts to complete the theory of quantum statistics by developing a theory of quantum dynamics~\cite{Bohm1952a,holland1995quantum,deBroglie1987,Pena2015} have all involved descriptions of particles guided by waves. The shortcomings of these dynamical theories have allowed some to retreat to the safe haven of the Copenhagen Interpretation, according to which the theory of quantum statistics developed in the 1930s is complete: there is no underlying particle dynamics. The recent discovery of a hydrodynamic pilot-wave system~\cite{couder2005walking} has provided a fresh perspective on the quantum problem, and a progressive approach to discerning what can and cannot be understood about quantum mechanics from a classical perspective~\cite{Bush2015a,BushOza2020}.

Pilot-wave theories have been developed to describe particle-wave interactions on both the microscopic and macroscopic scales. In quantum mechanics, the double-solution pilot-wave theory of Louis de Broglie ~\cite{deBroglie1923,deBroglie1930,deBroglie1987} was proposed in the 1920s on the premise that microscopic particles have an associated internal vibration at the Compton frequency that generates waves. The resonant wave-particle interaction was posited to result in the particle being propelled by its guiding or `pilot' wave with the de Broglie wavelength, giving rise to statistical behavior consistent with the standard formalism. While never developed to completion, de Broglie's theory provided a number of cornerstones of quantum theory, including the Einstein-de Broglie relation and the de Broglie relation, $p = \hbar k$. In 1952, David Bohm~\cite{Bohm1952a,Bohm1952b,holland1995quantum} proposed a pilot-wave theory that consists of a dynamic reformulation of the statistical theory of quantum mechanics, in which the particle is guided by its wavefunction, whose evolution is prescribed by the linear Schr\"odinger equation. According to Bohm's trajectory equation, the particle accelerates in response to classical and quantum potentials. The quantum potential, $Q$, is expressible in terms of the wavefunction, as prescribes the system's statistical behavior; thus, it is $Q$ that is responsible for the non-locality of Bohmian mechanics.

%Pilot-wave theory aims to characterize the underlying particle-wave interactions that dominate many experimental observations.  
Hydrodynamic pilot-wave theory~\cite{Bush2015a,Bush2015b,BushOza2020} has been developed to describe the motion of millimetric droplets self-propelling on the surface of a vibrating liquid bath, a system discovered by Yves Couder and Emmanuel Fort in 2005 ~\cite{couder2005walking,couder2006single}. By virtue of a resonant interaction between the bouncing droplets and the underlying wave field, the droplet is piloted by a quasi-monochromatic wavefield with the Faraday wavelength. This walking-droplet system is remarkable in that it represents a macroscopic realization of the type of pilot-wave dynamics envisaged by de Broglie, and all the more remarkable in that it has yielded a growing list of {\em hydrodynamic quantum analogs} (HQAs)~\cite{Bush2015a,Bush2018Chaos,BushOza2020}. These include analogs of single-particle diffraction and interference~\cite{couder_single-particle_2006,Pucci2016,Ellegaard2020}, orbital quantization~\cite{Fort2010,Harris2014a,Oza2014a,Perrard2014,Labousse2014a} tunneling~\cite{Eddi2009b,Nachbin2018,Tadrist2019}, quantum corrals~\cite{Harris2013a,Saenz2018}, Friedel oscillations~\cite{Saenz2019a}, spin lattices~\cite{Saenz2021} and long-range correlations in bipartite systems~\cite{Nachbin2018,Papatryfonos2021,NachbinPRF22}. In several of these systems, wave-like statistical forms emerge that are strikingly similar to those arising in their quantum counterparts~\cite{Harris2013a,harris2013pilot,Saenz2018,Saenz2019a}. HQAs thus suggest that a pilot-wave dynamics of the form engendered in the hydrodynamic system might plausibly underlie quantum statistics~\cite{Bush2015a,BushOza2020}. 

The feature of the walking-droplet system responsible for the emergent quantum features is `path-memory'~\cite{Eddi2011a}, as renders the drop dynamics non-Markovian, and results from the persistence of the pilot wave on the bath surface. The drop is propelled by its wave field, whose form is prescribed by the droplet's history and environment. The critical control parameter is the bath's vibrational acceleration, $\Gamma$, as prescribes the bath's proximity to the Faraday threshold, $\Gamma_F$, above which waves would form even in the absence of the droplet. While $\Gamma$ is always less than $\Gamma_F$ in the laboratory, the closer $\Gamma$ is to $\Gamma_F$, the more persistent the waves generated by the droplet, and the longer the path-memory. 
%The pilot-waves in such HQA systems are known to play a fundamental role in driving the dynamical evolution of the particle trajectory~\cite{harris2013pilot}.  Indeed, 
The manner in which the droplet dynamics changes as the vibrational acceleration (or `memory') is increased progressively towards the Faraday threshold has been characterized and rationalized for the free droplet~\cite{Molacek2013a,Molacek2013b,Oza2013}, droplet pairs~\cite{Arbelaiz2018,Oza2017,Couchman2019}, confined rings~\cite{Thomson2019,Thomson2020} and free rings~\cite{Couchman2020}. Transitions from steady orbital motion to chaotic dynamics have been reported and characterized in a number of settings, including motion in a rotating frame~\cite{Fort2010,Harris2014a,Oza2014a,Oza2014b}, motion confined by a central force~\cite{Perrard2014,Labousse2014a,Kurianski2017,durey2020bifurcations} and motion confined by boundaries~\cite{Harris2013a,Saenz2018,Cristea2018,Durey2018}.   Rich dynamical properties have been revealed, including period-doubling cascades to chaotic trajectories~\cite{Oza2014b,Harris2014a,tambasco2016onset,rahman2022walking}.  
The dynamical systems aspects of the walking-droplet system have been highlighted in the recent review of Rahman and Blackmore~\cite{rahman2020b}.

In characterizing the bifurcation structure of the walking-droplet system, prior work has focused primarily on the droplet dynamics,
specifically, how the particle trajectory changes with increasing memory. However, for confined walker motion, connections between the mean pilot-wave field and the emergent statistical forms have been both reported~\cite{Saenz2018} and rationalized~\cite{Durey2018,Durey2020_2}. As the resulting mean pilot-wave is expressible in terms of the droplet's statistical behavior, Bush \& Oza~\cite{BushOza2020} propose that it may play a role similar to that of the quantum potential in Bohmian mechanics. While the mean walker dynamics is thus effectively non-local, as is Bohmian mechanics, the time-resolved walker dynamics is entirely local.
The current work will be the first to focus on the bifurcation structure of the pilot wave as the system memory is increased progressively.  By leveraging the {\em dynamic mode decomposition} (DMD), the underlying pilot-wave field can be characterized, revealing its own period-doubling cascade to chaos.  The analysis provides a spatio-temporal modal analysis of the underlying low-dimensional wave interactions that characterize the overall pilot-wave physics. 

DMD originated as a modal analysis method in the fluid dynamics community.  Introduced as an algorithm by Schmid~\cite{schmid2008aps,schmid2010jfm}, it has rapidly become a commonly used data-driven analysis tool and the standard algorithm to approximate the Koopman operator from data~\cite{rowley2009jfm}.  Specifically, DMD was used to identify dominant spatio-temporal coherent fluid structures from high-dimensional time-series data.  The DMD analysis offered an alternative to standard dimensionality reduction methods such as the {\em proper orthogonal decomposition} (POD), which highlighted low-rank features in fluid flows using the computationally efficient {\em singular value decomposition} (SVD)~\cite{kutz:2013}.  The advantage of using DMD over SVD is that the DMD modes are linear combinations of the SVD modes that have a common linear (exponential) behavior in time, given by oscillations at a fixed frequency with growth or decay.  Specifically, optimized DMD~\cite{askham2017arxiv,sashidhar2021bagging} is a regression to solutions of the form
\begin{align}
\bx(t) = \sum_{j=1}^r b_j \bphi_j e^{\omega_j t}  ,
%= \bPhi \exp( \boldsymbol{\Omega} t)\bb,
\label{eq:DMDapprox}
\end{align}
where $\bx(t)$ is an $r$-rank approximation to a collection of state space measurements $\mathbf{x}_k=\mathbf{x}(t_k)$ $(k=1, 2, \cdots , n)$.  The algorithm finds the DMD eigenvalues $\omega_j$, DMD modes $\bphi_j$ and their loadings $b_j$.  The $\omega_j$ determines the temporal behavior of the system associated with a modal structure $\bphi_j$, thus giving a highly interpretable representation of the dynamics.  Such a regression can also be learned from time-series data~\cite{lange2020fourier}.  DMD may be thought of as a combination of the SVD/POD in space with the Fourier transform in time, combining the strengths of each approach~\cite{chen2012jns,kutz2016book}.  DMD is modular due to its simple formulation in terms of linear algebra, resulting in innovations related to control~\cite{proctor2016siads}, compression~\cite{brunton2015jcd,erichson2016arxiva,erichson2016jrtp}, reduced-order modeling~\cite{alla2017nonlinear}, and multi-resolution analysis~\cite{kutz2016multiresolution,champion2019discovery}, among others.

DMD also provides a natural mathematical connection with pilot-wave hydrodynamics.  Specifically, the emergent steady-state statistics of the walking droplets have been shown in certain instances to be comparable to solutions of the time-independent Schr\"odinger equation of quantum mechanics~\cite{Harris2013a,Saenz2018,BushOza2020}. The time-dependent Schr\"odinger equation takes the general form
\begin{equation}
  i\psi_t = -  \psi_{xx} + V(x) \psi    
\end{equation}
where $\psi(x,t)$ is the wavefunction, $V(x)$ is a potential and the coefficients have been normalized to unity.  The state-space variable ${\bf x}(t)$ (\ref{eq:DMDapprox}) is the vectorized version of the Schr\"odinger wavefield $\psi(x,t)$.  The general solution to the discretized
Schr\"odinger equation is given by the DMD approximation (\ref{eq:DMDapprox}).  In this case, the DMD modes and eigenvalues correspond to the eigenstates and their corresponding energy levels respectively of the quantum system.  Thus, DMD provides not only a valuable approximation method, but a regression to the solution form that is standard in quantum mechanics.

%\begin{figure}[t]
 %   \centering
    %\hspace*{-.6in}
  %  \begin{overpic}[width = .4\textwidth]{wave2.png}
   % \end{overpic}
%    \begin{overpic}[width = .4\textwidth]{wave3.png}
%    \end{overpic}
%   \caption{Canonical dynamics.}
   % \label{fig:pareto}
%    \label{fig:atmos_chem}
%\end{figure}

\begin{figure}[t]
    \centering
    %\hspace*{-.6in}
    \begin{overpic}[width = .25\textwidth,height=3.7cm]{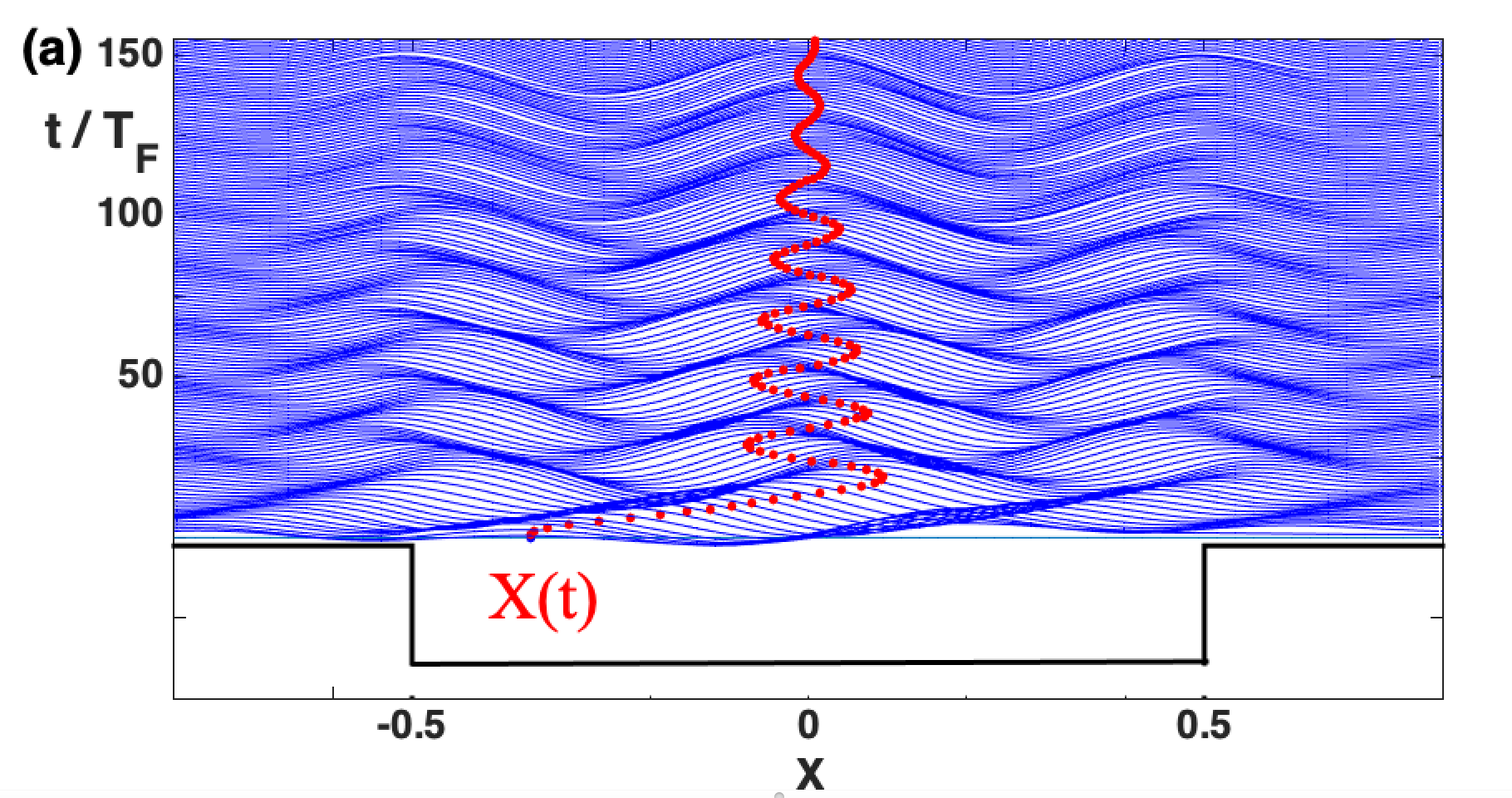}
    %\put(0,65){(a)}
    %\put(22,80){{\color{blue}{$\eta(x,t)$}}}
    %\put(40,30){{\color{red}{$X(t)$}}}
    \end{overpic}
    \hspace*{-.25in}
    \begin{overpic}[width = .25\textwidth,height=3.7cm]{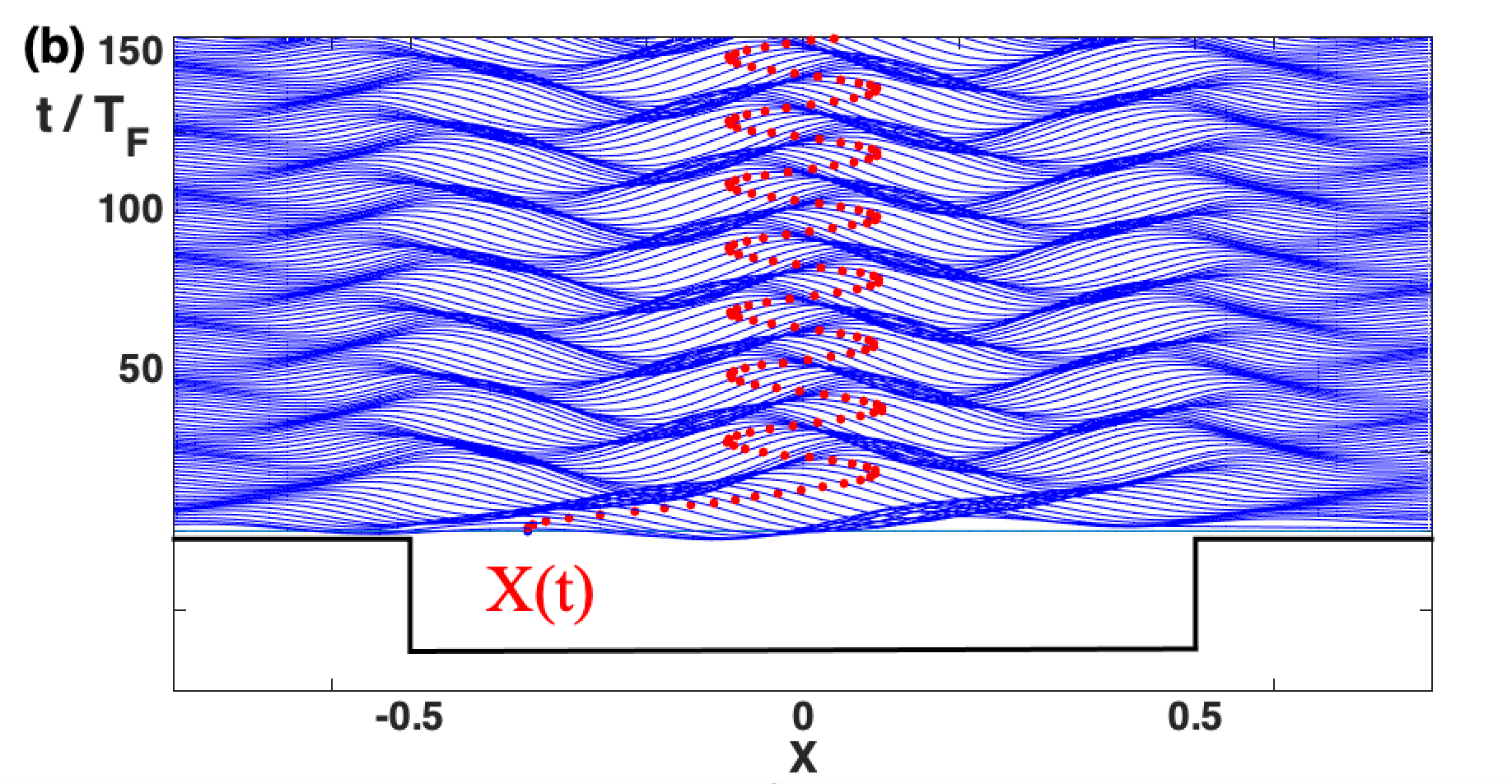}
    %\put(0,65){(b)}
    \end{overpic}
    \begin{overpic}[width = .25\textwidth,height=3.7cm]{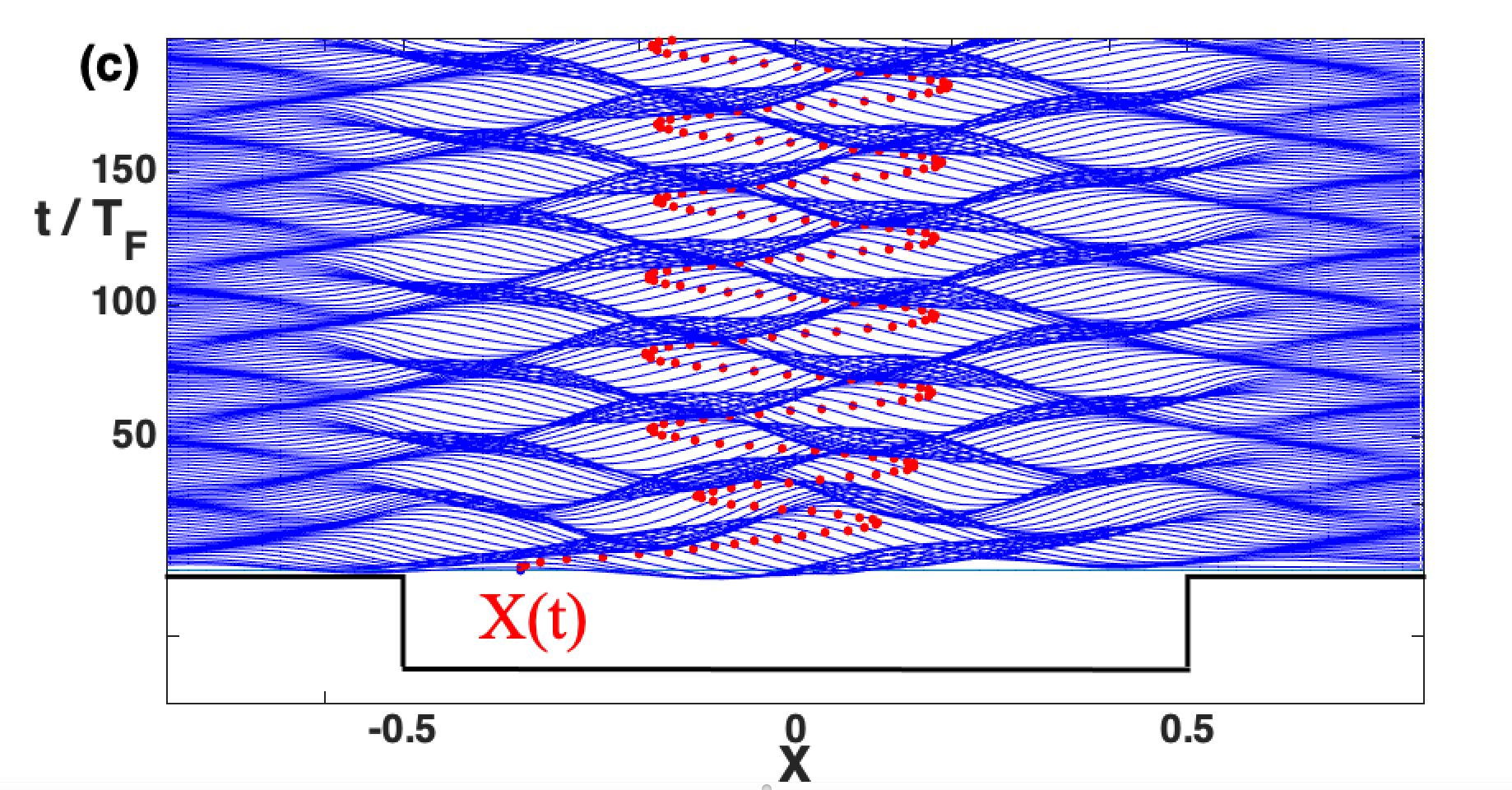}
    %\put(0,65){(c)}
    \end{overpic}
    \hspace*{-.25in}
    \begin{overpic}[width = .25\textwidth,height=3.7cm]{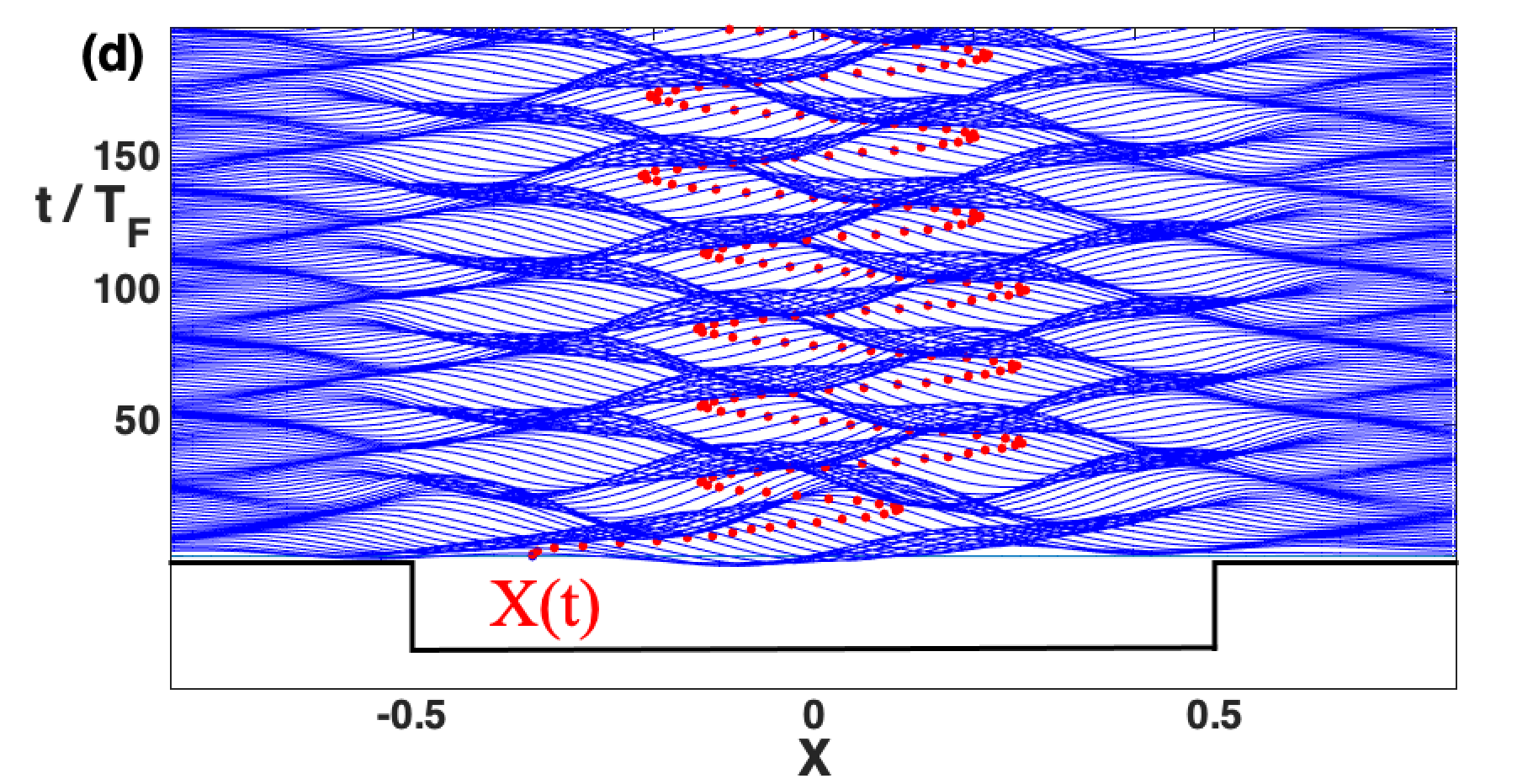}
    %\put(0,65){(d)}
    \end{overpic}
    \caption{Wave-droplet dynamics over 150 Faraday periods in a cavity of width 1cm. The vibrational forcing takes the values $\Gamma =$ (a) 4.8, (b) 5.0, (c) 5.3 and (d) 5.4. For this geometry the Faraday threshold is approximately $\Gamma_F$ = 6.15. All other parameters are kept unchanged, including the droplet's initial position at $X=-0.35$.  The wavefield $\eta(x,t)$ is denoted in blue while the horizontal droplet position $X(t)$ is indicated in red.}
   % \label{fig:pareto}
    \label{fig:wavesdroplet}
\end{figure}

%%%%%%%%%%%%%%%%%%%%%%%%%%%%%%%%%%%%%%%%%%%%%%%%%%%%%%%%%%%%%%%%%%
\section{PILOT WAVE PHYSICS}

The walking-droplet system is modeled theoretically through a trajectory equation for the droplet and a partial differential equation that describes the spatio-temporal dynamics of the accompanying waves~\cite{Molacek2013a,Molacek2013b,Oza2013}. The wave model may be 
obtained from an asymptotic simplification of the linear, free surface, Navier-Stokes equations \cite{JFM15}. 
The Helmholtz
decomposition of the velocity field yields, in the weakly viscous regime,  an irrotational velocity potential 
perturbed by a weak shearing component, arising through the streamfunction. The pilot wave along the undisturbed free surface $z=0$, is expressed through a weakly diffusive
Bernoulli law as well as a weakly diffusive kinematic condition, as detailed below. 
The velocity potential is denoted by $\phi(x,z,t)$.  
The velocity field components in the bulk of the fluid    %, associated to this harmonic function 
are given by $(u,v) = \nabla \phi$. 
The wave elevation is denoted by $\eta(x,t)$. The fluid parameters are $\rho$, the
fluid density, $\sigma$, the surface tension and  $\nu$, the kinematic viscosity.  
In the reference frame of the fluid bath, that oscillates at frequency $\omega_0$, gravity takes the form
$g(t) = g(1 + \Gamma \sin(\omega_0 t))$.
The free surface wave equations are given by    \cite{JFM15,Tunnel17}:
\begin{equation}
\frac{\partial \phi}{\partial t} (x,0,t) = - g(t) \eta + \frac{\sigma}{\rho} \eta_{xx}
% \frac{\partial^2\eta}{\partial x^2} 
+ 2\nu ~ \phi_{xx}
%\frac{\partial^2\phi}{\partial x^2}  
-  \!\! \frac{1}{\rho} 
P_d(x-X(t)),
\label{Bern}
\end{equation}
\begin{equation}
\frac{\partial \eta}{\partial t} (x,t) =  DtN[\phi] + 2\nu ~\eta_{xx}.
%\frac{\partial^2\eta}{\partial x^2}.
\label{Kin}
\end{equation}
The diffusive terms, in both the Bernoulli law and the kinematic condition, are the leading order terms 
from the vortical component of the Helmholtz decomposition. The presence of the droplet is felt through
the pressure term $P_d$, centered at the droplet's position $X(t)$, which acts as
a wave-maker.  This pressure term is compactly supported in space,
over the droplet's diameter, and is discontinuous in time, being activated periodically at each bounce. 
The velocity potential satisfies Laplace's equation which enables one to define  
the Dirichlet-to-Neumann (DtN) operator that maps the Dirichlet data $\phi(x,0,t)$
onto the free surface's normal speed, at time $t$:
\begin{equation}
DtN [\phi] =  \phi_z (x,0,t).
\label{DtN}
\end{equation}
The DtN operator is defined as a Fourier integral operator and is computed in a 
straightforward manner using a conformal mapping and the Fast Fourier Transform (FFT) \cite{Tunnel17}.

To complete the wave-particle model, the above wave system is coupled to the droplet's horizontal trajectory equation~\cite{Molacek2013b}:
\begin{equation}
m\frac{d^2X}{dt^2} + c~F(t)\frac{dX}{dt} = -F(t)\frac{\partial\eta}{\partial x}(X(t),t).
\label{DropODE}
\end{equation}
%This is basically Newton's law, where the forcing term comes from the wave dynamics.
%The slope of the wave elevation $\eta$ provides the forcing on the droplet, indicating how the pilot-wave guides the particle.
The drop is propelled by the wave force, as is proportional to the local gradient of the wave field, and resisted by a linear drag.
The  magnitude of the propulsive wave force transmitted
during the contact time, here prescribed as $Tc=T_F/4$ where $T_F$ is the Faraday period, 
is denoted by $F(t)$ . 
This time-dependent coefficient also appears in the drag term, since drag is also imparted during impact.
Further modeling details can be found elsewhere~\cite{JFM15, Tunnel17}.

%\begin{figure}[t]
%    \centering
%    \hspace*{-.3in}
%    \begin{overpic}[width = %.5\textwidth]{droplet.eps}
%    \end{overpic}
%    \caption{Canonical dynamics.}
%   % \label{fig:pareto}
%    \label{fig:PDE_sims}
%\end{figure}

\begin{figure}[t]
    \centering
    %\hspace*{-.6in}
    \begin{overpic}[width = .23\textwidth]{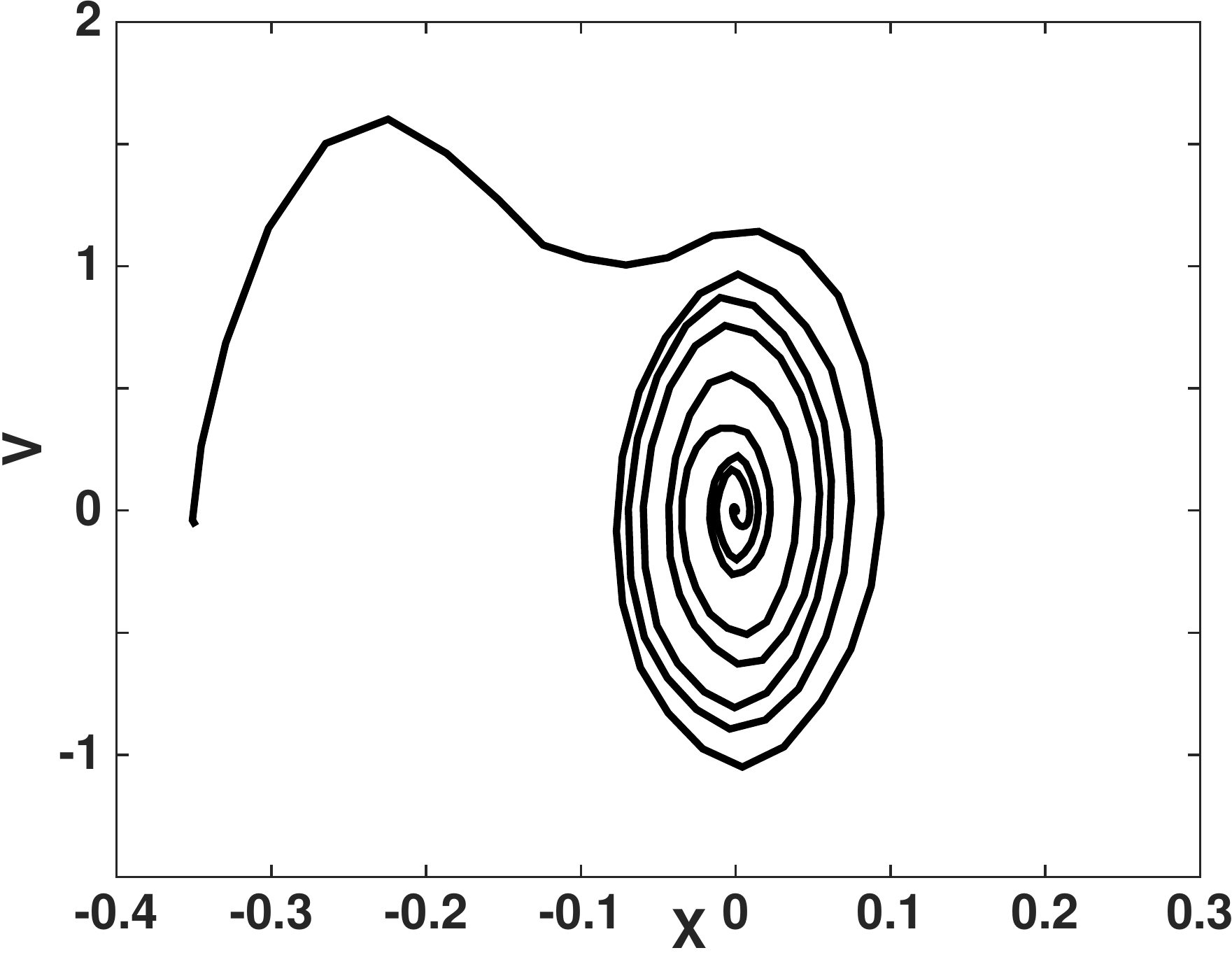}
    \end{overpic}
    \begin{overpic}[width = .23\textwidth]{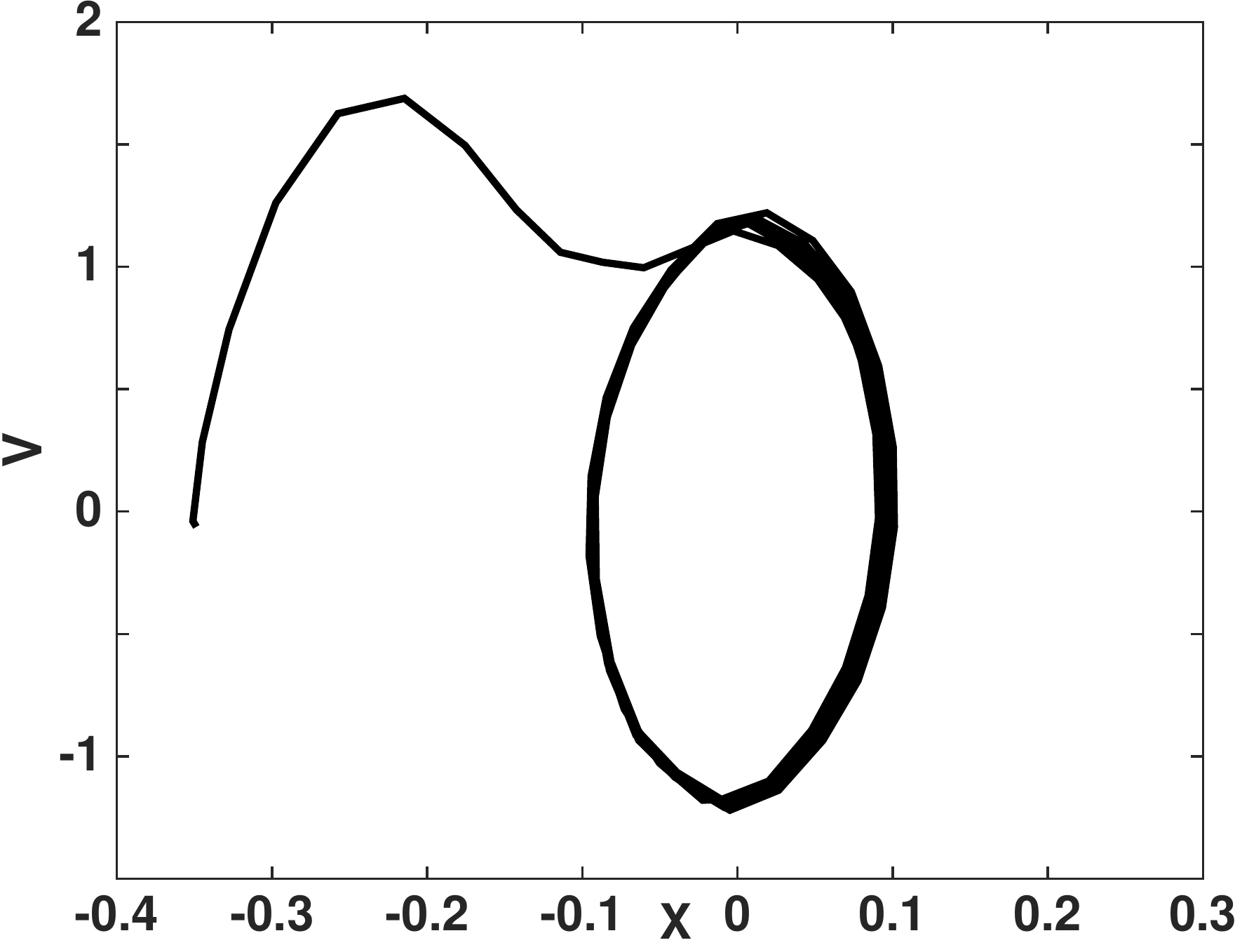}
    \end{overpic}\\
    \begin{overpic}[width = .23\textwidth]{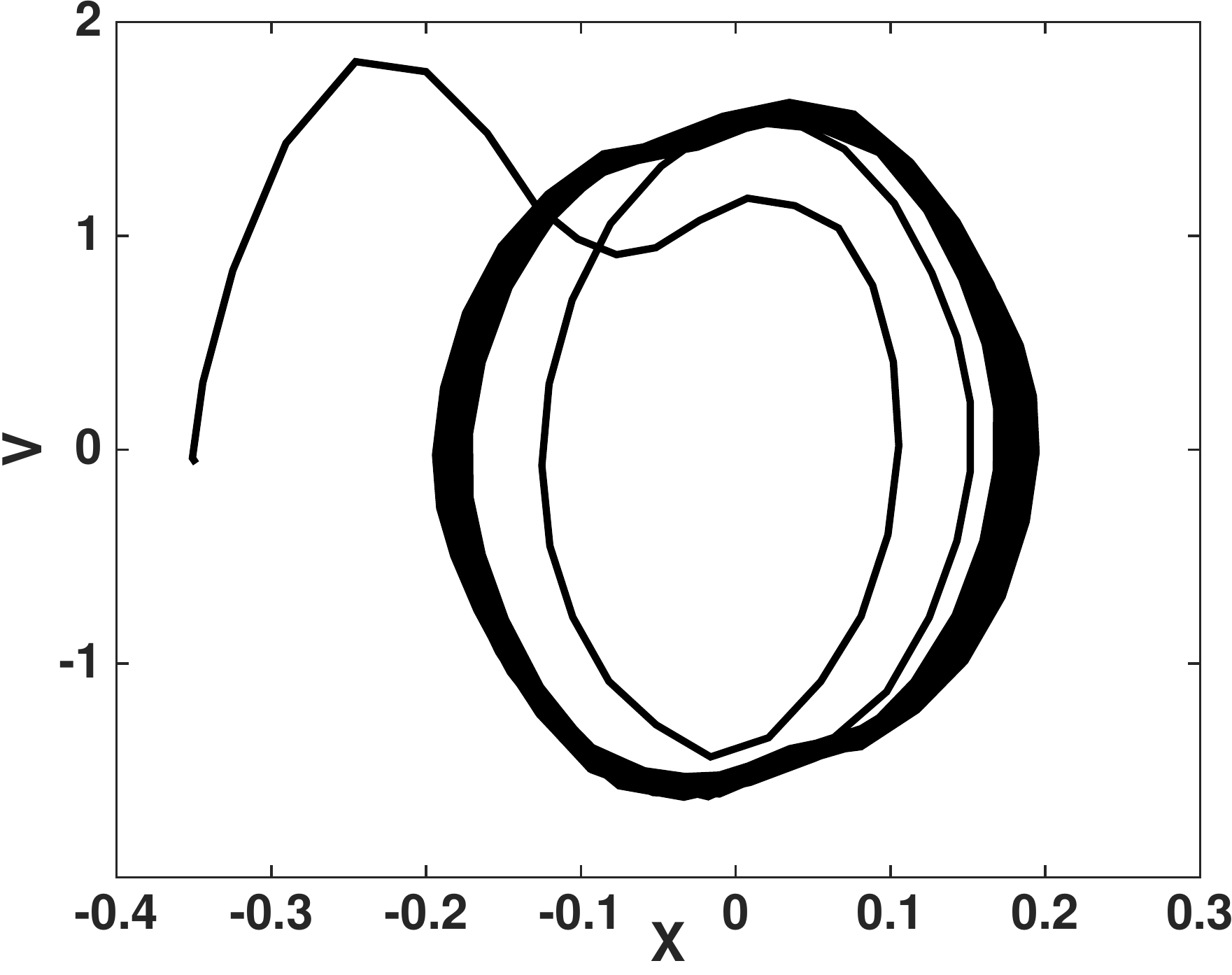}
    \end{overpic}
    \begin{overpic}[width = .23\textwidth]{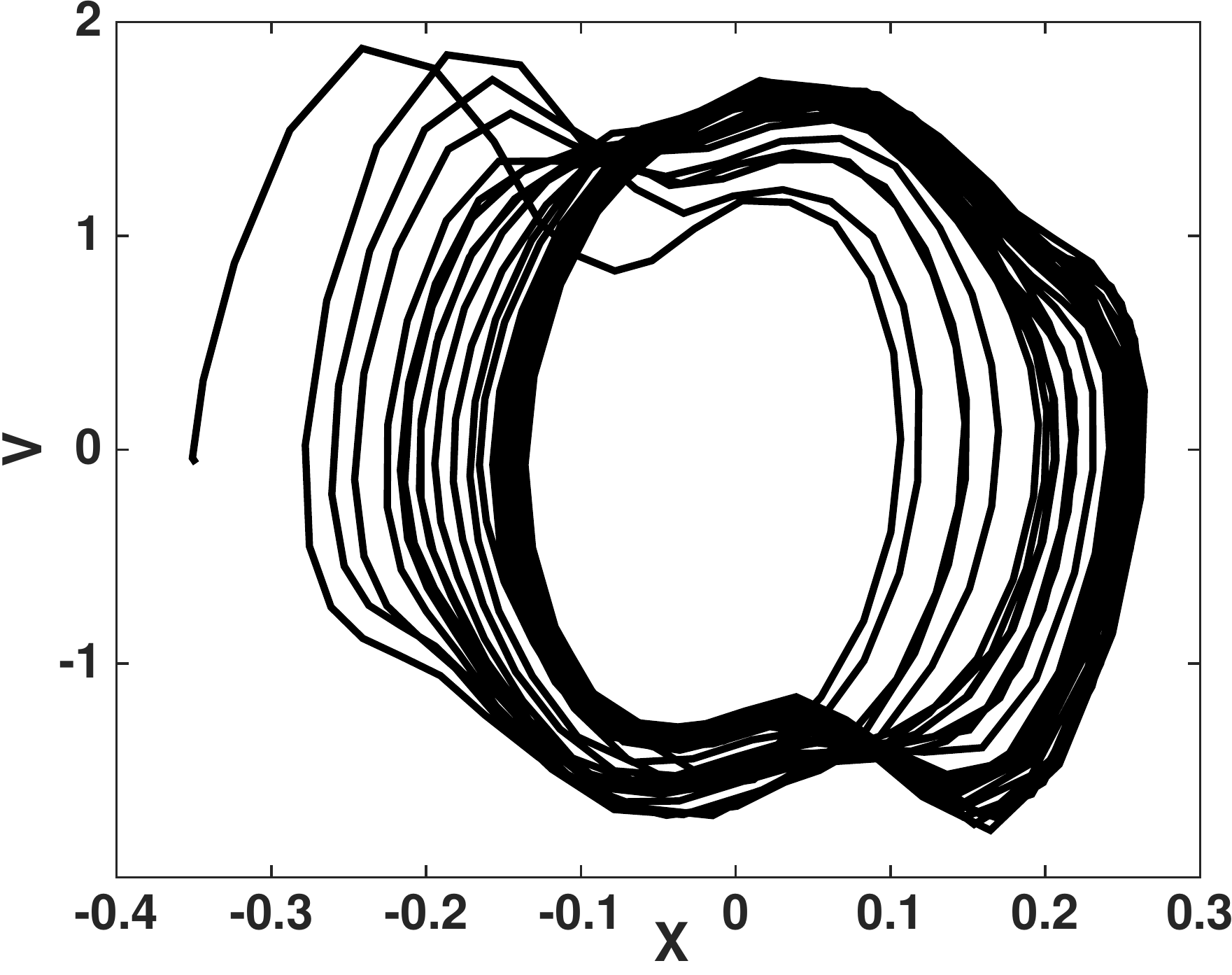}
    \end{overpic}
    \caption{Phase space dynamics. The droplet position is denoted by $X$ and the droplet speed by $V=\dot{X}$. Panels correspond to extensions of the sequences illustrated in Fig. \ref{fig:wavesdroplet}, for which $\Gamma$ = 4.8, 5.0, 5.3 and 5.4, and the initial particle position $X(0)=-0.35$. The total time for the simulations is $t=4000 T_F$ for (a-b) and $t=8000T_F$ for (c-d). }
%    to better highlight variations in the oscillator's range.}
    \label{fig:PDE_sims}
\end{figure}

\begin{figure}[t]
    \centering
    %\hspace*{-.6in}
    \begin{overpic}[width = .45\textwidth]{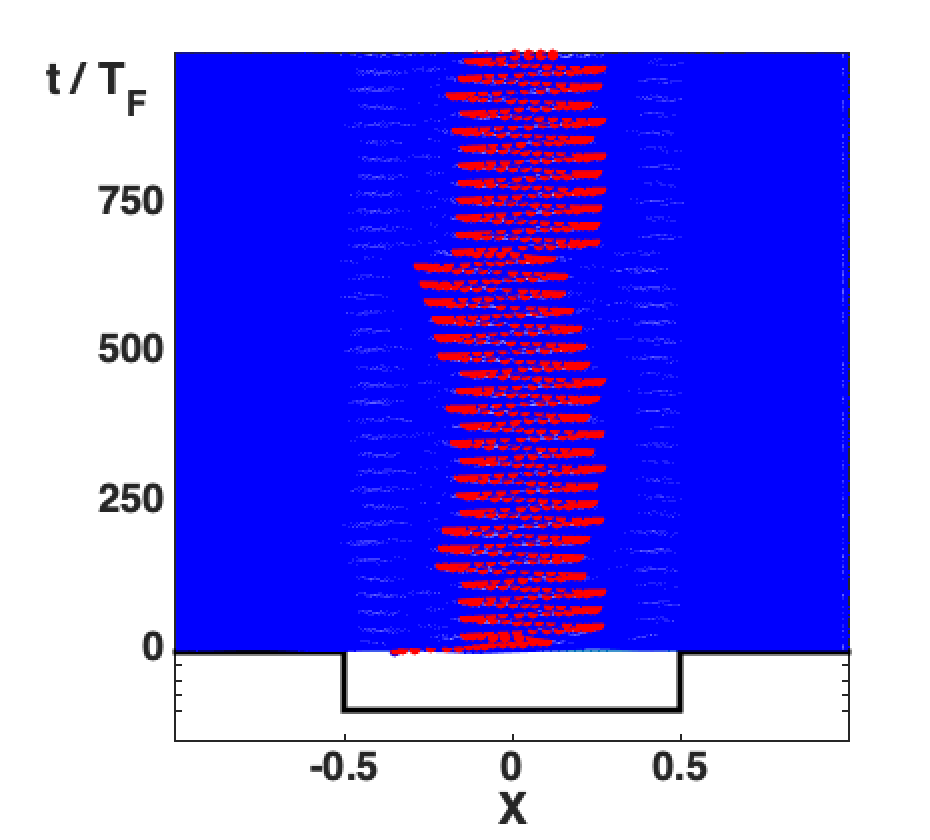}
    \put(25,88){{\color{blue}{$\eta(x,t)$}}}
    \put(50,88){{\color{red}{$X(t)$}}}
    \end{overpic}
    \caption{Simulation of particle-wave dynamics with $\Gamma=5.4$, displayed over a longer time interval
    than in Fig.~\ref{fig:wavesdroplet}d.  The wavefield $\eta(x,t)$ is denoted in blue while the droplet position $X(t)$ is indicated in red.}
    \label{fig:longer wavesdroplet}
\end{figure}

\begin{figure}[t]
    \centering
    \hspace*{-.2in}
    \begin{overpic}[width = .55\textwidth]{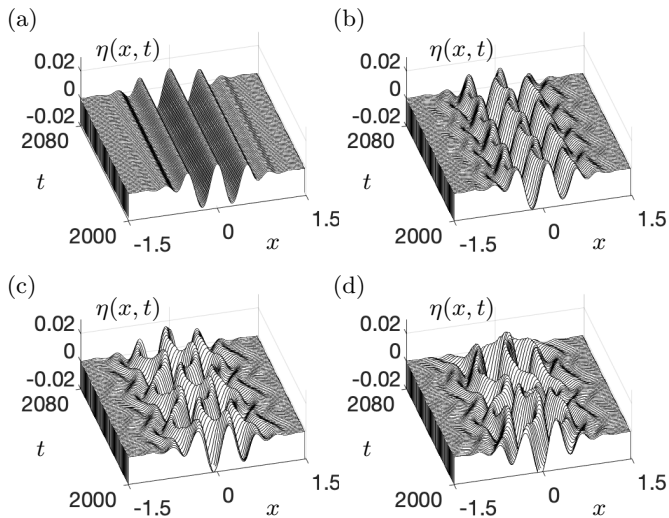}
    \put(3,70){(a)}
    \put(47,70){(b)}
    \put(38,4){$x$}
    \put(82,4){$x$}
    \put(38,40){$x$}
    \put(7,48){$t$}
    \put(82,40){$x$}
    \put(51,48){$t$}
    \put(51,12){$t$}
    \put(7,12){$t$}
    \put(3,34){(c)}
    \put(47,34){(d)}
    \put(15,66){{$\eta(x,t)$}}
    \put(60,66){{${\eta}(x,t)$}}
    \put(15,31){{$\eta(x,t)$}}
    \put(60,31){{${\eta}(x,t)$}}
    \end{overpic}
    \caption{Pilot-wave dynamics corresponding to Fig.~\ref{fig:wavesdroplet}, where $\Gamma$ takes on the values (a) 4.8, (b) 5.0, (c) 5.3 and (d) 5.4, respectively.  The pilot wave field is shown for $t\in[2000,2080]$ after the droplet behavior has settled onto its long time dynamics, as illustrated in Figs.~\ref{fig:wavesdroplet} and \ref{fig:PDE_sims}.}
   % \label{fig:pareto}
    \label{fig:wave0}
\end{figure}

\begin{figure}[t]
    \centering
    \hspace*{-.0in}
    \begin{overpic}[width = .45\textwidth]{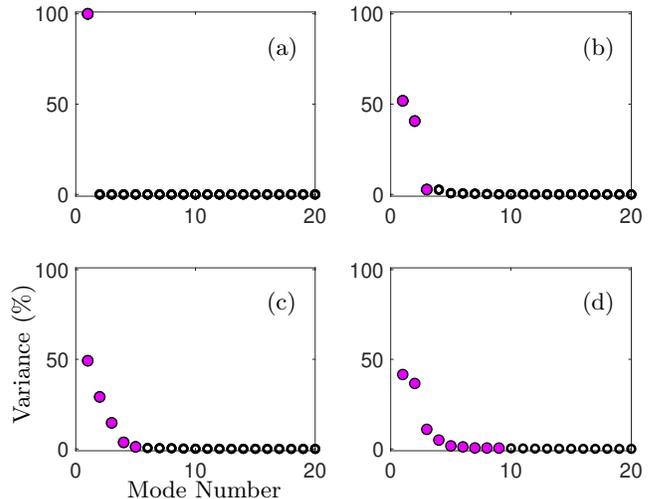}
        \put(38,70){(a)}
    \put(90,70){(b)}
    \put(38,28){(c)}
    \put(90,28){(d)}
    \put(15,-3){Mode Number}
    \put(-4,9){\rotatebox{90}{Variance ($\%$)}}
    %\put(-4,8){\rotatebox{90}{$100\,\sigma_j/\sum \sigma_k$}}
    \end{overpic}
    \caption{Low-rank structure of the pilot-wave dynamics corresponding to Fig.~\ref{fig:wave0}, where $\Gamma$ takes on the values (a) 4.8, (b) 5.0, (c) 5.3 and (d) 5.4. The singular value decomposition shows the percentage of variance in each SVD mode, i.e. the $j$th mode variance is equal to $100\sigma_j/\sum \sigma_k$.  The variance of only the first 20 modes are shown, with the magenta modes depicting the number of modes required for accurate DMD reconstruction.}
   % \label{fig:pareto}
    \label{fig:svd_cut}
\end{figure}

The evolution of the PDE model is shown in Fig.~\ref{fig:wavesdroplet}. As the forcing amplitude is increased, the walking-droplet system undergoes a bifurcation sequence.  For low-amplitude forcing, $\Gamma = 4.8$, the droplet bounces in place in the middle of the well (Fig.~\ref{fig:wavesdroplet}a).  As the forcing amplitude is increased, the droplet begins to oscillate periodically in the well  (Fig.~\ref{fig:wavesdroplet}b).  Further increase of $\Gamma$ generates larger oscillations until eventually period doubling occurs, which is detailed in Sec.~\ref{sec:dynamics}.
The dynamics of the droplet can also be plotted in phase-space using the droplet position, denoted by $X$, and the droplet speed, denoted by $V=\dot{X}$.  Figure~\ref{fig:PDE_sims} shows the underlying attractors associate with the dynamics illustrated in Fig.~\ref{fig:wavesdroplet}.
%as a function of increasing forcing $\Gamma$.
%
Increasing $\Gamma$ leads to a
chaotic motion of the droplet, as further detailed in Fig.\ref{fig:longer wavesdroplet}. The center of oscillation varies in a chaotic fashion.  As will be detailed in Sec.~\ref{sec:dynamics}, the bifurcation sequence exhibits a period-doubling bifurcation sequence to chaos, which is canonical for damped, driven systems~\cite{li2010geometrical,spaulding2002nonlinear,ding2009operating,koch2021multiscale}.

The corresponding pilot-wave dynamics of Fig.~\ref{fig:PDE_sims} is highlighted in Fig.~\ref{fig:wave0}. 
%In what follows, it will be shown that  as $\Gamma$ is increased, a    
%Thus the pilot-wave counterpart of the droplet physics is the observation of 
%a period doubling route to chaos is observed.
%in the spatio-temporal fluid field.  
As $\Gamma$ takes on the values 4.8, 5.0, 5.3 and 5.4, the pilot wave evolves from a steady field, to a periodic wave to a doubly periodic wave to spatio-temporal chaos.  In what follows, we will illustrate that the spatio-temporal dynamics can be completely characterized by the DMD decomposition.  Indeed, DMD exploits the low-rank structure of the wave field as shown in Fig.~\ref{fig:svd_cut}.
%for $\Gamma$ taking on the values 4.8, 5.0, 5.3 and 5.4.

%%%%%%%%%%%%%%%%%%%%%%%%%%%%%%%%%%%%%%%%%%%%%%%%%%%%%%%%%%%%%%%%%%
\section{DYNAMIC MODE DECOMPOSITION}

The DMD algorithm can be best understood from the so-called {\em exact} DMD~\cite{tu2014jcd}, which is simply a least-square fitting procedure.  Specifically, the exact DMD algorithm seeks a best fit linear operator $\mathbf{A}$ that approximately advances the state of a system, $\mathbf{x}\in\mathbb{R}^n$, forward in time according to the linear dynamical system
\begin{align}\label{Eq:DMD:Propagator}
    \mathbf{x}_{k+1} = \mathbf{A}\mathbf{x}_k,
\end{align}
where $\mathbf{x}_k=\mathbf{x}(k\Delta t)$, and $\Delta t$ denotes a fixed time step that is small enough to resolve the highest frequencies in the dynamics.  Thus, the operator $\mathbf{A}$ is an approximation of the Koopman operator $\mathcal{K}$ restricted to a measurement subspace spanned by direct measurements of the state $\mathbf{x}$~\cite{rowley2009jfm,brunton2021modern}.

Bagheri~\cite{bagheri2014} first highlighted that DMD is particularly sensitive to the effects of noisy data, with systematic biases introduced to the eigenvalue distribution~\cite{duke2012error,bagheri2013jfm,dawson2016ef,hemati2017tcfd}.  As a result, a number of methods have been introduced to stabilize performance, including total least-squares DMD~\cite{hemati2017tcfd}, forward-backward DMD~\cite{dawson2016ef}, variational DMD~\cite{azencot2019consistent}, subspace DMD~\cite{takeishi2017}, time-delay embedded DMD~\cite{brunton2017natcomm,arbabi2017,kamb2020siads,hirsh2021structured}, robust DMD methods~\cite{askham2017robust,scherl2020prf}, and 
physics-informed DMD~\cite{baddoo2021physics}.
However, the {\em optimized DMD} algorithm of Askham and Kutz~\cite{askham2017arxiv}, which uses a variable projection method for nonlinear least squares to compute the DMD for unevenly timed samples, provides the best and optimal performance of any algorithm currently available. This is because it directly solves the exponential fitting problem of DMD. 
Consider the data matrix
\begin{equation}
\bX = \begin{bmatrix} \vline & \vline & & \vline \\
\bx(t_1) & \bx(t_2) & \cdots & \bx(t_m) \\
 \vline & \vline & & \vline
 \end{bmatrix} .
\end{equation}
The optimized DMD algorithm directly solves the exponential fitting problem in order to produce the the matrix decomposition
Thus, the data matrix $\mathbf{X}$ may be reconstructed as
\begin{eqnarray}
      \bX &\approx&  \bPhi \mbox{diag}(\bb) {\bf T}(\boldsymbol{\omega})
  \nonumber \\
  &=& \left[ \! \begin{array}{ccc} | & & | \\ \boldsymbol{\phi}_1 & 
    \!\!\cdots\!\! & \boldsymbol{\phi}_r \\ | & & | \end{array} \! \right] \!\!
    \left[ \! \begin{array}{ccc} b_1 &  & \\ & \!\!\ddots\!\! & \\ & & b_r  \end{array} \!\right] \!\!
    \left[ \! \begin{array}{ccc} e^{\omega_1 t_1} & \!\cdots\! & e^{\omega_1 t_m} \\
    \vdots & \!\ddots\! & \vdots \\ e^{\omega_r t_1} & \!\cdots\! & e^{\omega_r t_m} \end{array} \! \right]
    .
\label{eq:dmd_opt}
\end{eqnarray}
The variable projection method for this approximation~\cite{askham2017arxiv} determines the parameters of the matrix components.  Thus the optimization is formulated as 
\begin{equation}
\mbox{argmin}_{ \boldsymbol{\omega}, \bPhi_{\bf b} } \| \bX -   \bPhi_{\bf b} {\bf T}(\boldsymbol{\omega}) \|_F,
\end{equation}
where $\bPhi_b$ = $\bPhi \bf{diag(b)}$. Thus optimized DMD provides a direct approximation to the solution (\ref{eq:DMDapprox}), which is a regression to the form of the modal solution expansion to the vectorized Schr\"odinger equation. This has been shown to provide a superior decomposition due to its ability to optimally suppress bias and handle snapshots collected at arbitrary times.  
Moreover, with statistical bagging, the BOP-DMD algorithm (bagging, optimized DMD) also provides robust models with uncertainty quantification in the presence of noise, corruption and outliers~\cite{sashidhar2021bagging}.  Physically motivated constraints can also be embedded into the DMD architecture~\cite{baddoo2021physics}, further enhancing the algorithm.
The disadvantage of optimized DMD is that one must solve a nonlinear optimization problem, which can fail to converge.

\section{Inferring particle statistics from wave dynamics}

Durey {\it et al.}~\cite{Durey2018} deduced a simple relation between the mean pilot-wave field and the droplet statistics for walker motion in a statistically steady state, either periodic or ergodic. For a particle moving in a bounded region (for example, when constrained by a central force), the mean wave field $\overline{\eta}$ may be 
expressed as the convolution of the wave field of a stationary bouncer $\eta_B$ and the droplet's histogram, $\mu({\bf x})$, normalized to have total mass one. Note that in this instance, where boundary effects are negligible, $\eta_B$ is independent of the droplet position but does depend on the system memory.
For a domain with 
variable bottom topography relevant for the cavity considered here, the result was generalized 
by Durey {\it et al.}~\cite{Durey2020_2} through incorporation of the spatial dependence of the bouncer wave field
resulting from the influence of boundaries.
The convolution is thus generalized to an integral operator, where the kernel 
$\eta_G(x,y)$ is given by the wave field of a bouncer located at position $y$. In this case,
the statistics of the particle and its pilot wave are related by
%The relation between the particle statistics and the mean wave field may thus be expressed as 
\cite[Eq. 3.2]{Durey2020_2}:
\begin{equation}
\overline{\eta}(x) = \int_{-\infty}^\infty \eta_G(x,y) \mu(y) \, d y
\label{Eq:durey}
\end{equation}
where $\overline{\eta}(x) = \lim_{N\rightarrow \infty} \frac{1}{N} \sum_{k=1}^N \eta(x,t_k)$,
is the time-averaged wave field and $t_k = kT_F$.
%equally spaced in some time interval $[0,T]$}. 
% and $\mu(x)$ is the particle's histogram. and $\eta_G(x,y)$ is the wave field produced by a stationary droplet bouncing at $y$.}
We note that numerical verification of (\ref{Eq:durey}) was not presented in \cite{Durey2020_2}. In
panel (e) of Figures~\ref{fig:wave1}-\ref{fig:wave4} we compare the time-averaged wave field, computed directly from the numerical simulations, 
with that predicted by expression (\ref{Eq:durey}). The agreement is excellent, even for the cases where the phase-space orbits indicate complex cycles.

We proceed by exploring relation (\ref{Eq:durey}) in a novel fashion. 
Inverting the integral operator allows us to infer particle-statistics from the mean wave field (as may be viewed as an effective potential) 
computed over the time interval of interest. The particle statistics is 
thus obtained with no specific knowledge of the particle dynamics, 
a step that evokes quantum mechanics. Numerical aspects of the operator inversion are presented in the
Appendix.
%{\bf NK: I added an overline in ETA in ALL panels (f). In all captions the order (e) and (f) are inverted. I think it is best to invert the figures (e) and (f) since average comes before inversion.}
Panel (f) in Figures~\ref{fig:wave1}-\ref{fig:wave4} compare the particle's probability density function, as deduced directly from particle tracking, with that produced by inverting equation (\ref{Eq:durey}). 
The agreement between the two is satisfactory in all cases.
Thus, without having recorded the particle dynamics,  we obtain a good approximation to the particle statistics from 
the time-averaged wave field $\overline{\eta}$.

%Panels (e) and (f) in Figures~\ref{fig:wave1}-\ref{fig:wave4} illustrate the particle histograms as calculated directly by particle tracking, and via the convolution result (\ref{Eq:durey}). 
 
\section{DYNAMICS AND BIFURCATIONS}
\label{sec:dynamics}

The optimized DMD algorithm is used on the wave data shown in Fig.~\ref{fig:wave0} using a low-rank truncation suggested by Fig.~\ref{fig:svd_cut}.  Specifically, as the 
forcing parameter $\Gamma$ takes on the values 4.8, 5.0, 5.3 and 5.4. the low-rank structure is well-captured by one, three, five and nine modes respectively.  As will be detailed below, the number of modes used was dictated by the minimum number of modes required to reproduce an accurate representation of the dynamics.

\begin{figure}[t]
    \centering
    \hspace*{-.1in}
    \begin{overpic}[width = .55\textwidth]{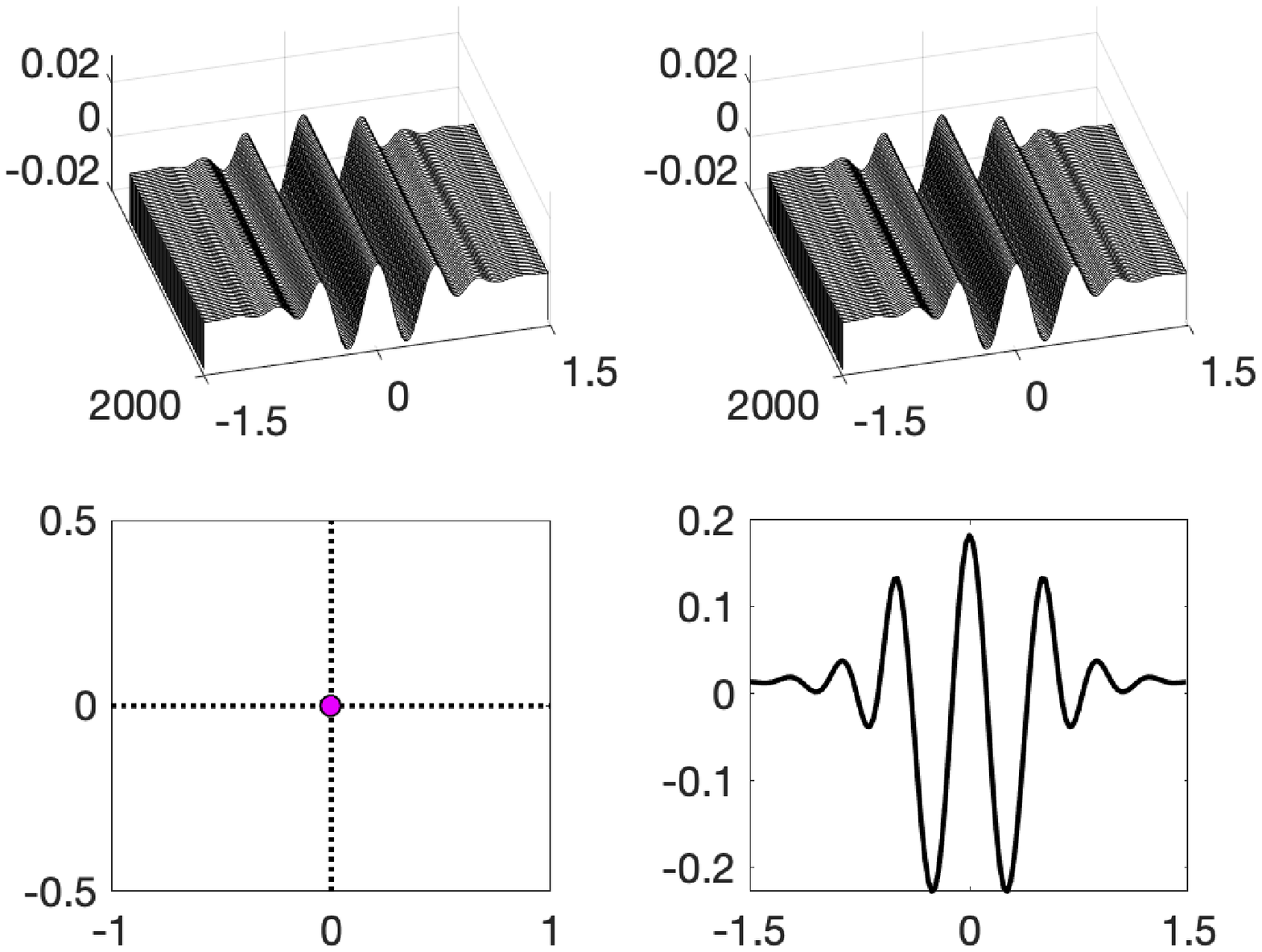}
    \put(3,70){(a)}
    \put(47,70){(b)}
    \put(80,4){$x$}
    \put(38,40){$x$}
    \put(7,48){$t$}
    \put(82,40){$x$}
    \put(51,48){$t$}
    \put(3,30){(c)}
    \put(47,30){(d)}
    \put(78,30){$\phi_1(x)$}
    \put(23,36){$\Im\{ \omega_1 \}$}
    \put(33,17){$\Re\{ \omega_1 \}$}
    \put(15,66){{$\eta(x,t)$}}
    \put(60,66){{$\tilde{\eta}(x,t)$}}
    \end{overpic}
        \hspace*{0.2in}
    \begin{overpic}[width = .44\textwidth]{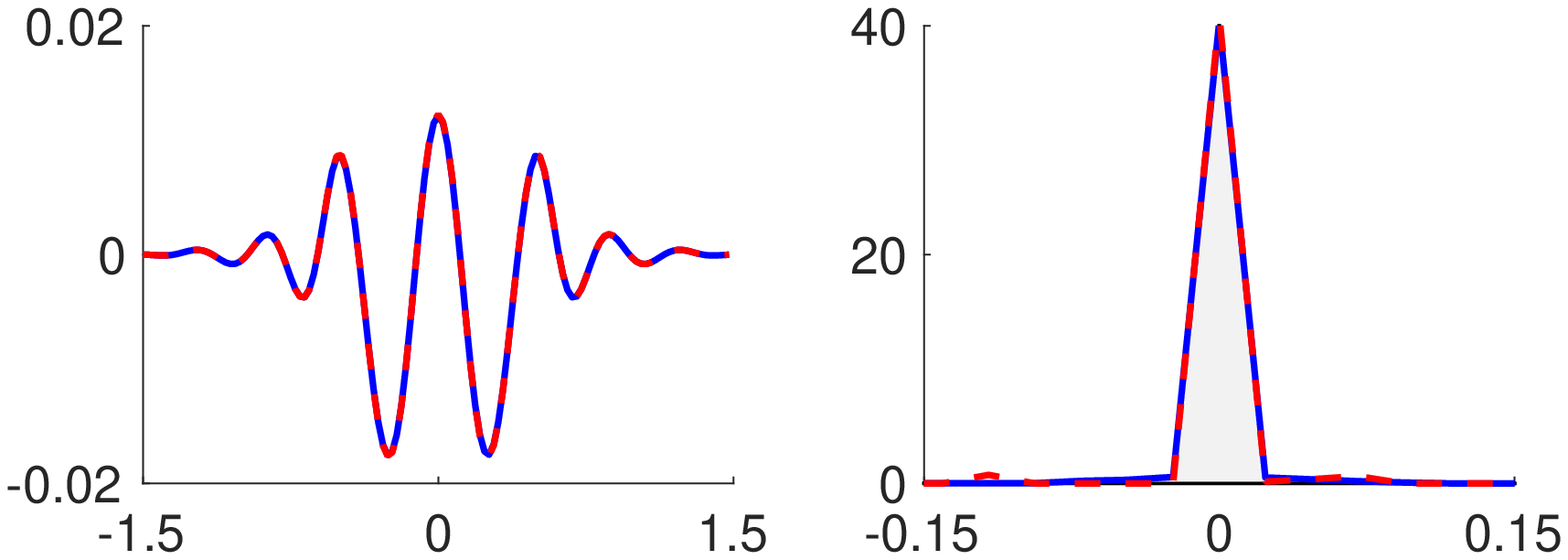}
        \put(-6,30){(e)}
    \put(49,30){(f)}
    \put(85,30){$\mu(x)$}
    \put(35,30){$\overline{\eta}(x)$}
    \put(35,-1){$x$}
    \put(90,-1){$x$}
    \end{overpic}
    \caption{Pilot-wave dynamics for the lowest forcing value considered, $\Gamma=4.8$.  
    The simulation was run for 4000 Faraday periods (or droplet bounces).
    Figure~\ref{fig:svd_cut} shows that a rank one decomposition is sufficient to model the steady-state wave dynamics.  In panel (a), the full PDE evolution is shown while in panel (b) the one-mode DMD approximation is shown.  Panel (c) shows the DMD eigenvalue while panel (d) shows the DMD eigenfunction used for reconstruction of the spatio-temporal dynamics.  Note that the DMD eigenvalue is at the origin which is consistent with the steady-state behavior in this parameter regime.  (e) The time-averaged pilot-wave field calculated directly (solid blue line) and deduced from the measured particle probability distribution via (\ref{Eq:durey}) (dashed red line).  (f) The particle probability distribution as computed via particle tracking (solid blue line) and inferred from the mean pilot wave by inverting equation (\ref{Eq:durey}) (dashed red line).  }
   % \label{fig:pareto}
    \label{fig:wave1}
\end{figure}

%\begin{figure}[t]
%    \centering
%    %\hspace*{-.3in}
%    \begin{overpic}[width = .5\textwidth]{bif1.eps}
%    \end{overpic}\\[.2in]
%  \begin{overpic}[width = .5\textwidth]{bif2.eps}
%    \end{overpic}\\[.2in]
%    \begin{overpic}[width = .5\textwidth]{bif3.eps}
%    \end{overpic}
%    \caption{Canonical dynamics.}
%   % \label{fig:pareto}
%    \label{fig:atmos_chem}
%\end{figure}

\begin{figure}[t]
    \centering
    \hspace*{-.1in}
    \begin{overpic}[width = .55\textwidth]{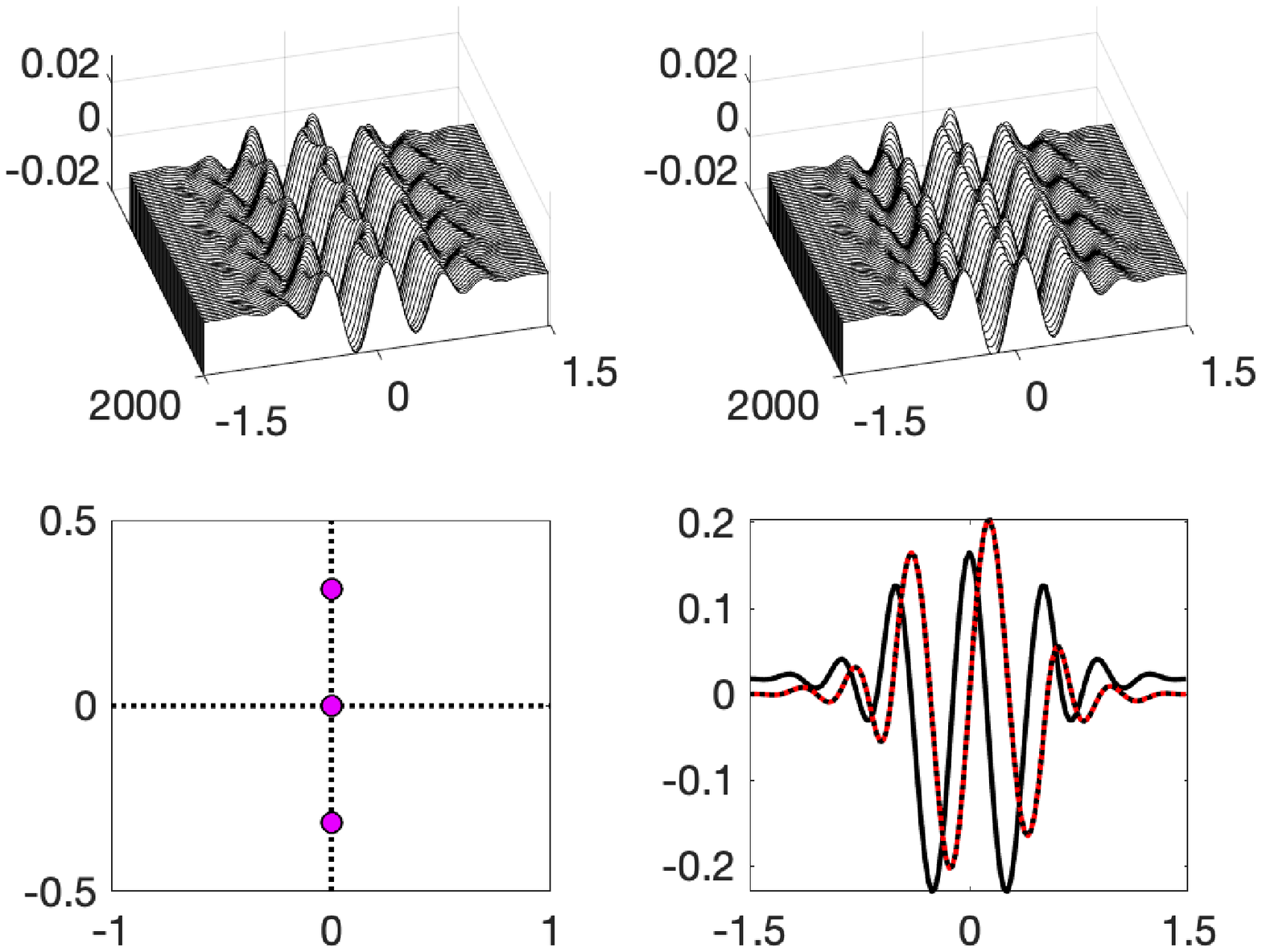}
    \put(3,70){(a)}
    \put(47,70){(b)}
    \put(80,4){$x$}
    \put(38,40){$x$}
    \put(7,48){$t$}
    \put(82,40){$x$}
    \put(51,48){$t$}
    \put(3,30){(c)}
    \put(47,30){(d)}
    \put(77,30){$\phi_k(x)$}
    \put(23,36){$\Re\{ \omega_n \}$}
    \put(33,17){$\Im\{ \omega_n \}$}
    \put(15,66){{$\eta(x,t)$}}
    \put(60,66){{$\tilde{\eta}(x,t)$}}
    \end{overpic}
            \hspace*{0.2in}
    \begin{overpic}[width = .44\textwidth]{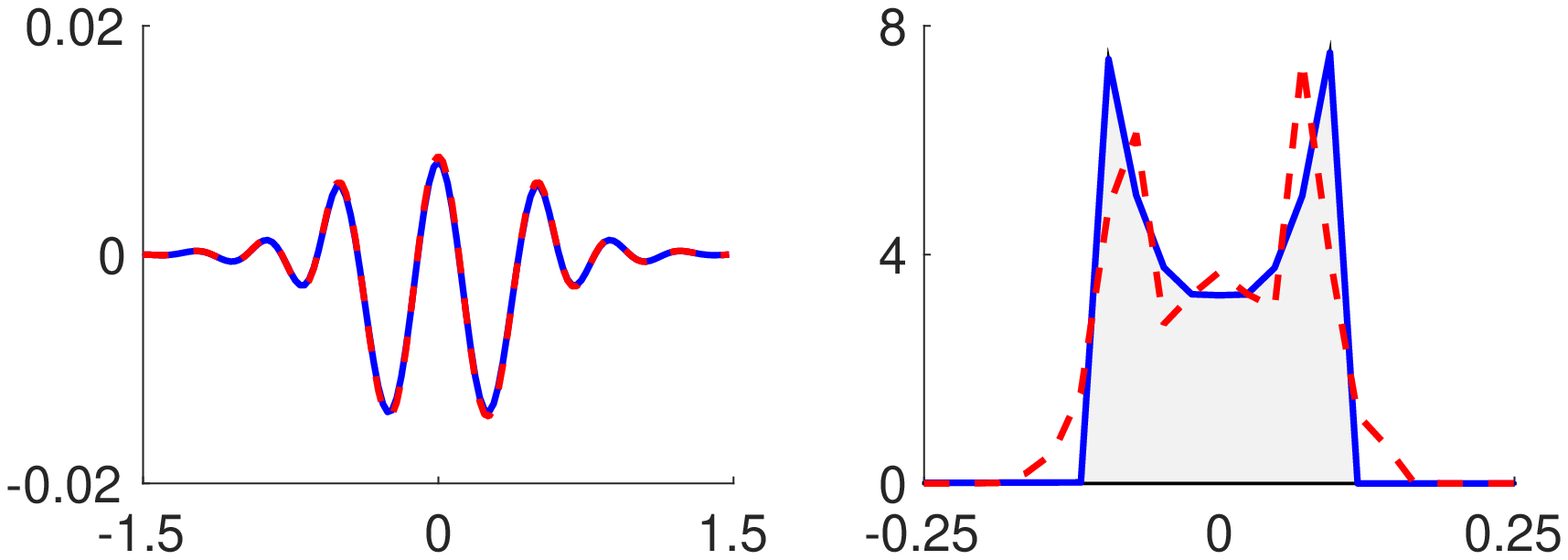}
        \put(-6,30){(e)}
    \put(49,30){(f)}
    \put(88,30){$\mu(x)$}
    \put(35,30){$\overline{\eta}(x)$}
    \put(35,-1){$x$}
    \put(90,-1){$x$}
    \end{overpic}
    \caption{Pilot-wave dynamics for $\Gamma=5.0$.  
     The simulation was run for 4000 Faraday periods.
     Figure~\ref{fig:svd_cut} shows that a rank three decomposition is adequate to model the spatio-temporal periodic behavior observed.  In panel (a), the full PDE evolution is shown while in panel (b) the three-mode DMD approximation is shown.  Panel (c) shows the DMD eigenvalues while panel (d) shows the DMD eigenfunctions used for reconstruction of the spatio-temporal dynamics. (The black line shows the steady-state background mode of Fig.6).  Note that the DMD eigenvalues are manifest in complex conjugate pairings typical of an underlying Hopf bifurcation.  (e) The time-averaged pilot-wave field (solid blue line) and that deduced from the measured particle probability distribution via (\ref{Eq:durey}) (dashed red line).  (f) The particle probability distribution as computed via particle tracking (solid blue line) and inferred from the mean pilot wave 
    by inverting equation (\ref{Eq:durey}) (dashed red line).  }
   % \label{fig:pareto}
    \label{fig:wave2}
\end{figure}

\begin{figure}[t]
    \centering
    \hspace*{-.1in}
    \begin{overpic}[width = .55\textwidth]{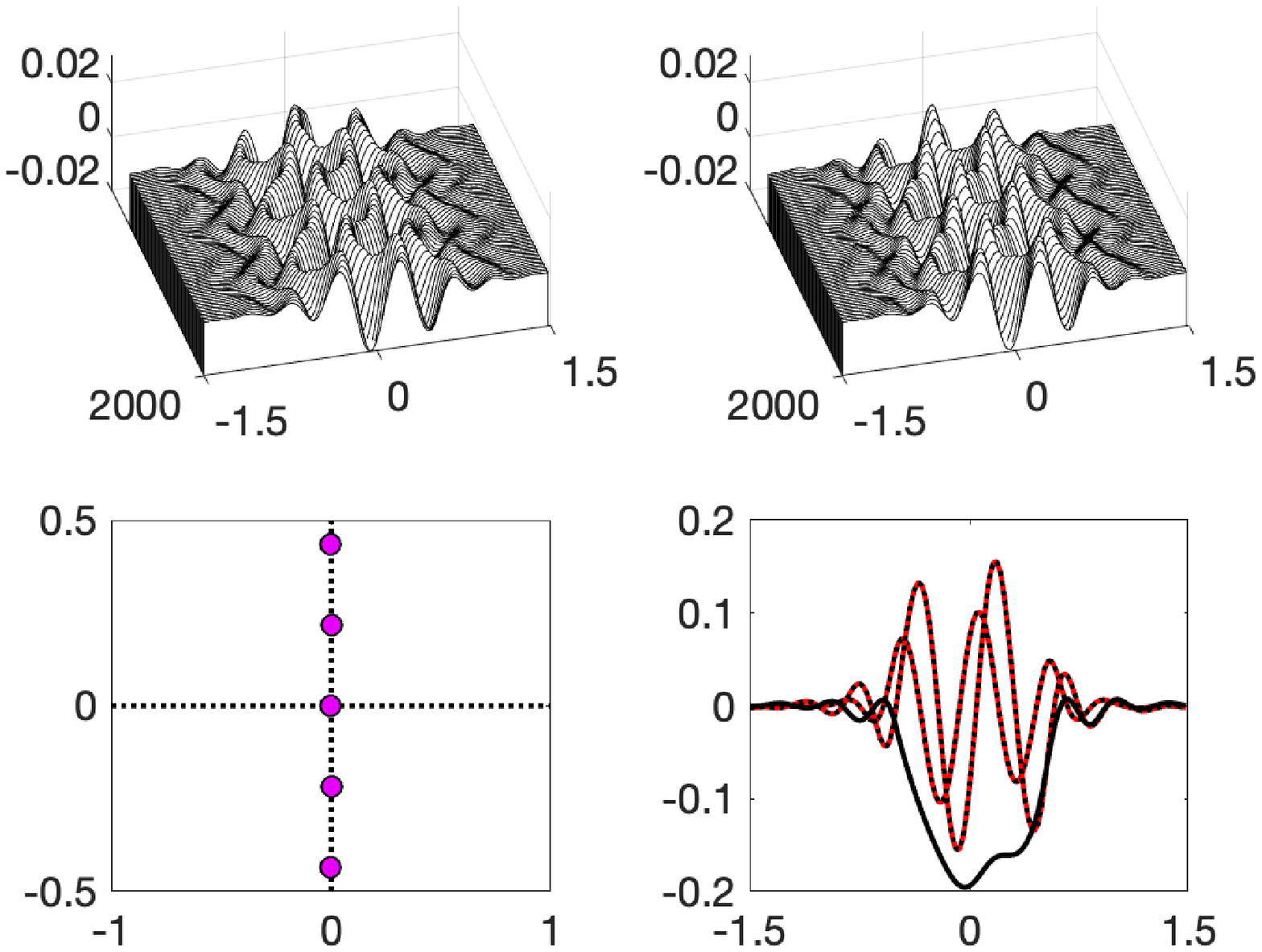}
    \put(3,70){(a)}
    \put(47,70){(b)}
    \put(80,4){$x$}
    \put(38,40){$x$}
    \put(7,48){$t$}
    \put(82,40){$x$}
    \put(51,48){$t$}
    \put(3,30){(c)}
    \put(47,30){(d)}
    \put(77,30){$\phi_k(x)$}
    \put(23,36){$\Re\{ \omega_n \}$}
    \put(33,17){$\Im\{ \omega_n \}$}
    \put(15,66){{$\eta(x,t)$}}
    \put(60,66){{$\tilde{\eta}(x,t)$}}
    \end{overpic}
            \hspace*{0.2in}
    \begin{overpic}[width = .44\textwidth]{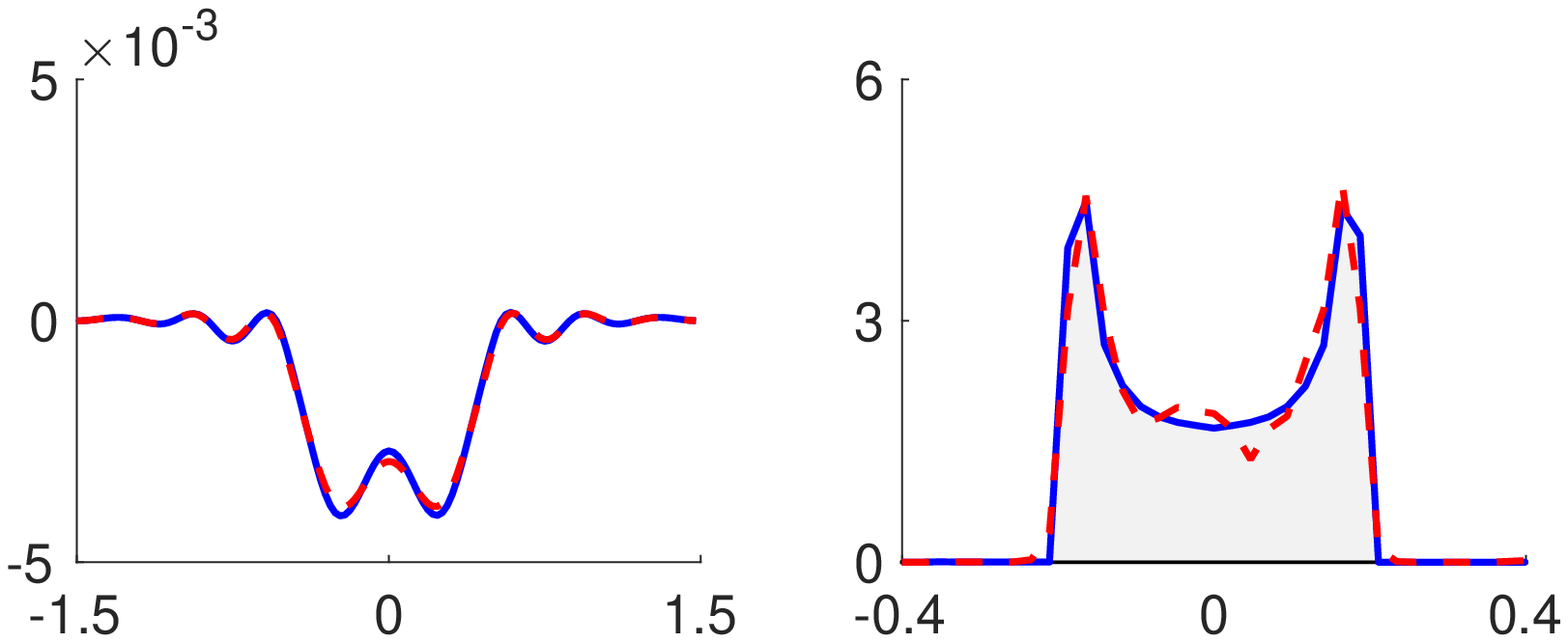}
        \put(-6,30){(e)}
    \put(49,30){(f)}
    \put(85,30){$\mu(x)$}
    \put(35,30){$\overline{\eta}(x)$}
    \put(35,-1){$x$}
    \put(90,-1){$x$}
    \end{overpic}
    \caption{Pilot-wave dynamics for $\Gamma=5.3$.  
     The simulation was run of 8000 Faraday periods.
     Figure~\ref{fig:svd_cut} shows that a rank five decomposition is adequate to model the spatio-temporal periodic behavior observed.  In panel (a), the full PDE evolution is shown while in panel (b) the five-mode DMD approximation is shown.  Panel (c) shows the DMD eigenvalues while panel (d) shows the DMD eigenfunctions used for reconstruction of the spatio-temporal wave dynamics. (The black line shows the steady-state background mode).  Note that the DMD eigenvalues appear in complex conjugate pairs whose frequencies are approximately harmonics, leading to the secondary period-doubling dynamics manifest in the pilot wave field. (e) The time-averaged pilot-wave field as computed directly (solid blue line) and deduced from the particle probability distribution via (\ref{Eq:durey}) (dashed red line).  (f) The particle probability distribution as computed via particle tracking (solid blue line) and inferred from the mean pilot wave 
    by inverting equation (\ref{Eq:durey}) (dashed red line). } 
   % \label{fig:pareto}
    \label{fig:wave3}
\end{figure}

\begin{figure}[t!]
    \centering
    \hspace*{-.1in}
    \begin{overpic}[width = .55\textwidth]{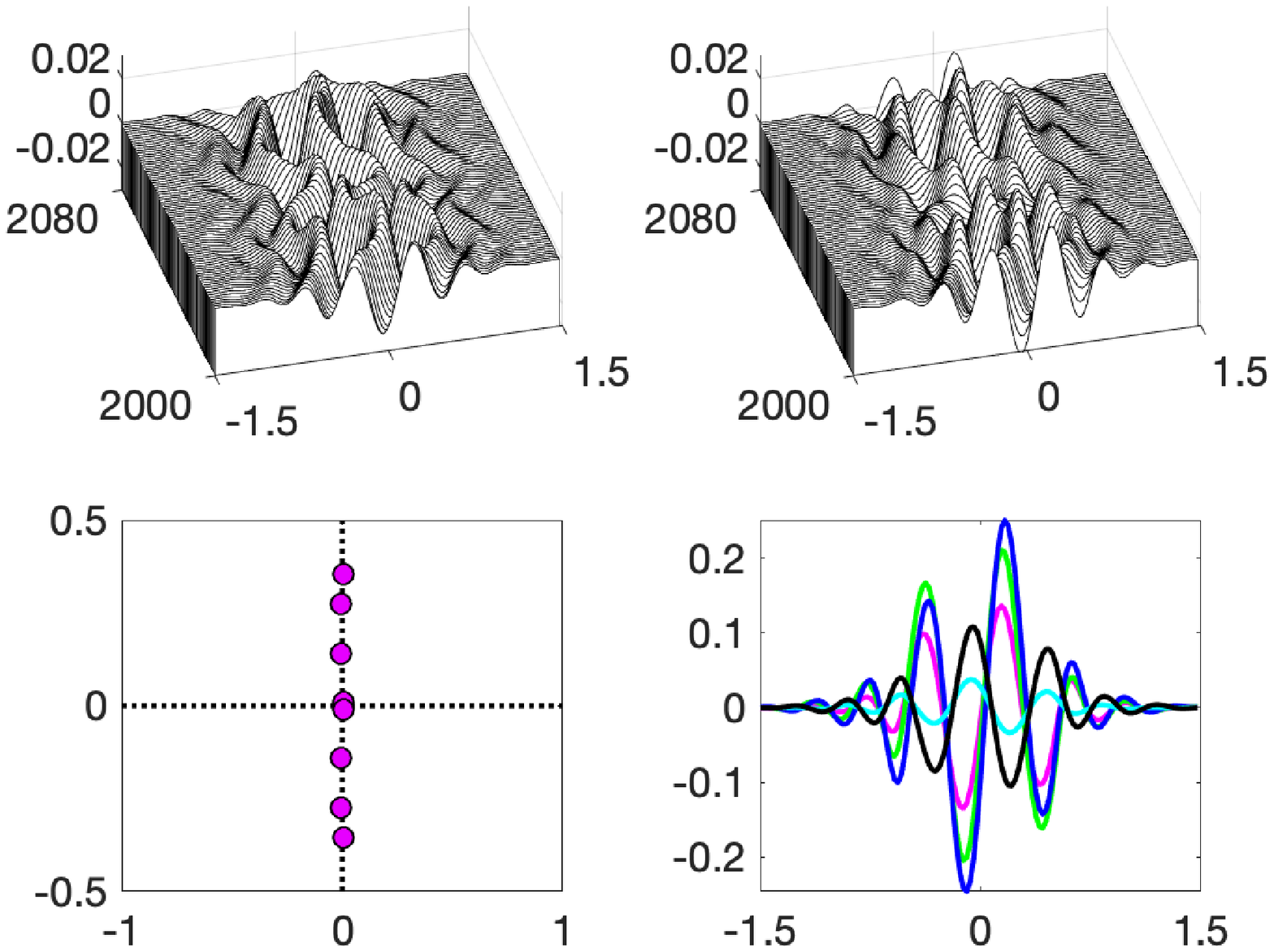}
    \put(3,70){(a)}
    \put(47,70){(b)}
    \put(80,4){$x$}
    \put(38,40){$x$}
    \put(7,48){$t$}
    \put(82,40){$x$}
    \put(51,48){$t$}
    \put(3,30){(c)}
    \put(47,30){(d)}
    \put(77,30){$\phi_k(x)$}
    \put(23,36){$\Re\{ \omega_n \}$}
    \put(33,17){$\Im\{ \omega_n \}$}
    \put(15,66){{$\eta(x,t)$}}
    \put(60,66){{$\tilde{\eta}(x,t)$}}
    \end{overpic}
            \hspace*{0.2in}
    \begin{overpic}[width = .44\textwidth]{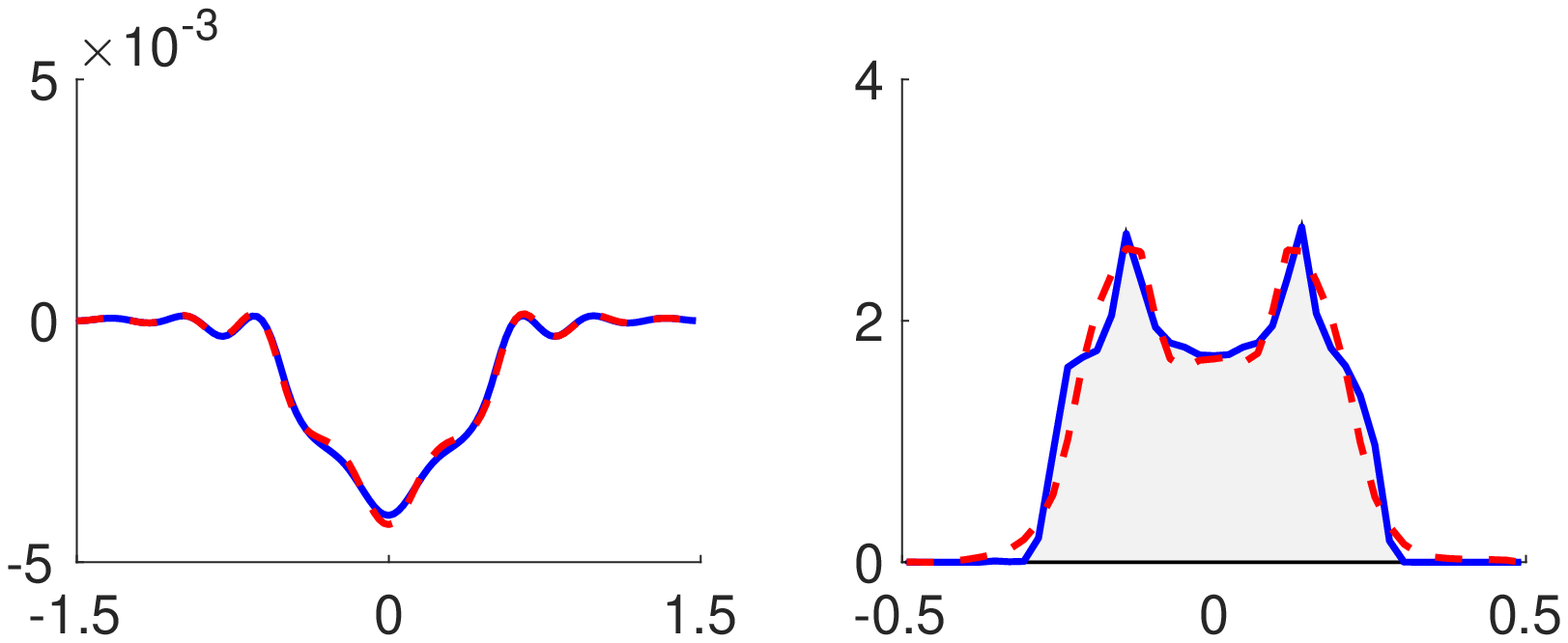}
        \put(-6,30){(e)}
    \put(49,30){(f)}
    \put(85,30){$\mu(x)$}
    \put(35,30){$\overline{\eta}(x)$}
    \put(35,-1){$x$}
    \put(90,-1){$x$}
    \end{overpic}
    \caption{Pilot-wave dynamics for the highest forcing value considered, $\Gamma=5.4$. 
    The simulation was run for 8000 Faraday periods.
    Figure~\ref{fig:svd_cut} shows that a rank nine decomposition is adequate to model the spatio-temporal chaotic behavior observed.  In panel (a), the full PDE evolution is shown while in panel (b) the nine-mode DMD approximation is shown.  Panel (c) shows the DMD eigenvalues while panel (d) shows the DMD eigenfunctions used for reconstruction of the spatio-temporal wave dynamics. (The black line shows the steady-state background mode as computed over the limited time interval shown).  Note that the DMD eigenvalues appear in complex conjugate pairings whose frequencies are incommensurate, which is responsible for the relatively complex dynamics.  (e) The time-averaged pilot-wave field computed directly (solid blue line) and that deduced from the measured particle probability distribution via (\ref{Eq:durey}) (dashed red line).  (f) The particle probability distribution as computed via particle tracking (solid blue line) and inferred from the mean pilot wave by inverting equation (\ref{Eq:durey}) (dashed red line).  }
   % \label{fig:pareto}
    \label{fig:wave4}
\end{figure}

Figures~\ref{fig:wave1}-\ref{fig:wave4} illustrate the dynamics as the dynamics transitions from the lowest value of $\Gamma=4.8$ to the highest value of $\Gamma=5.4$.  
Each of the four figures have six panels which illustate the characteristics dynamics as the forcing parameter $\Gamma$ takes on the values 4.8, 5.0, 5.3 and 5.4.  Panel (a) shows the wave dynamics after the decay of all transients in the system (Fig.~\ref{fig:PDE_sims}).  Recall that the wave-field has been strobed to remove oscillations at the Faraday frequency.  Panel (a) shows the evolution of the full PDE simulations, while panel (b) shows the low-rank DMD reconstruction.  The accompanying DMD eigenvalue and DMD eigenfunction are shown in panels (c) and (d).   Note that the eigenvalue at the origin captures a stationary behavior while the complex conjugate pairs of  eigenfunctions capture the periodic dynamics of the wavefield. Panel (e) shows the mean pilot-wave field (solid blue line) and that deduced from Durey’s convolution theorem~\cite{Durey2018,Durey2020_2} (red dashed line).  Panel (f) shows the particle probability distribution as computed directly from  particle tracking (solid blue line) and inferred from the mean pilot wave by inverting equation (\ref{Eq:durey}) (dashed red line).

Note that as the bifurcation sequence progresses, a period doubling bifurcation is observed with the generation of harmonics in the DMD spectra (See Figs~\ref{fig:wave2}-\ref{fig:wave3}).  Once the forcing is sufficiently large, the complex conjugate pairs are no longer harmonics, thus producing the observed spatio-temporal chaos. The overall period doubling route to chaos is thus evident in the simulations.

\begin{table*}[t!]
\label{tab:1}
\caption{Comparison of the quantum description of a particle in a one-dimensional well, and the pilot-wave system considered here.}
%\centering\renewcommand\cellalign{cc},
%\setcellgapes{3pt}\makegapedcells
\begin{tabular}{|c|c|c|} \hline
 & \textbf{Quantum mechanics} & \textbf{Pilot-wave hydrodynamics} \\ \hline 
\textbf{Driving parameter} & Particle energy & Memory parameter $\gamma$ \\ \hline
\textbf{Waveform} & Quantum wave function $\Psi$ & Faraday pilot wave $h(x,t)$\\ \hline
\textbf{Statistical inference} & Born's Rule &  Integral Operator Inversion~\cite{Durey2020_2}\\
\hline\end{tabular}
\end{table*}

The evolution from Figs.~\ref{fig:wave1} to \ref{fig:wave4} shows the underlying onset of instability in this pilot-wave hydrodynamic system.  This period-doubling is canonical for damped, driven systems, from optics to combustion~\cite{li2010geometrical,spaulding2002nonlinear,ding2009operating,koch2021multiscale}.  The optimized DMD algorithm provides a data-driven algorithm that provides a regression to the exponential solution form~(\ref{eq:DMDapprox}) commonly used to model physical systems, from quantum mechanics to electrodynamics.  Thus, DMD provides interpretable models and a clear quantification of the evolution of the pilot-wave field accompanying the droplet dynamics.

% %%%%%%%%%%%%%%%%%%%%%%%%%%%%%%%%%%%%%%%%%%%%%%%%%%%%%%%%%%%%%%%%%%
% \section{DROPLETS ON PILOT WAVES}    

% Model of droplet motion:  generalization of van der Pol equation:
% %
% \begin{equation}
%     x'' + \mu \exp(-\beta x^2)\left( 4x^4/3-4 x^2 +1 \right) x' + x = 0 
% \end{equation}
% %
% Dynamics are in Fig.~\ref{fig:generalvdp}.  In terms of Hermite polynomials, this is
% %
% \begin{equation}
%     x'' + \mu \exp(-\beta x^2) H_4(x) x' + x = 0 
% \end{equation}
% %
% Or alternatively,
% %
% \begin{equation}
%     x'' + \mu \psi_4(x) x' + x = 0 
% \end{equation}
% %
% where $\psi_n(x)$ is the Gauss-Hermite polynomials.

% \begin{figure}[t]
%     \centering
%     \hspace*{-.3in}
%     \begin{overpic}[width = .40\textwidth]{vdp_dyn.eps}
%     \end{overpic}
%     \caption{Generalization of van der Pol dynamics where the nonlinear damping is now related to the pilot-wave shape.}
%   % \label{fig:pareto}
%     \label{fig:generalvdp}
% \end{figure}

% \begin{figure}[t]
%     \centering
%     \hspace*{-.3in}
%     \begin{overpic}[width = .5\textwidth]{oscillate.eps}
%     \end{overpic}
%     \caption{Generalization of van der Pol dynamics where the nonlinear damping is now related to the pilot-wave shape.}
%   % \label{fig:pareto}
%     \label{fig:generalvdp}
% \end{figure}

%[NK - PLEASE ADD PANELS ef TO FIGS 6-9. AND I GUESS WE SHOULD DISCUSS THEM
%MORE EXTENSIVELY HERE.]

\section{CONCLUSIONS}

Particle-wave interactions arise throughout the physical sciences.  The specific example considered here of the pilot-wave hydrodynamic system has generated significant interest due to its connection to 
%the statistical nature of the theory of 
quantum mechanics.  Indeed, pilot-wave hydordynamics has provided a compelling example of well-resolved, classical particle-wave interactions producing quantum-like statistics.
To date, the particle dynamics has received the bulk of the mathematical attention, and the emergent statistics have been
seen as a compelling feature of the system that is not always simply rationalized.  
%In this case, the dynamics of a bouncing droplet, which is self-propelled along the surface of a vibrating bath, can produce rich dynamical behaviors and emergent statistics reminiscent of quantum systems.  
Here, we have instead focused on the evolution of the pilot-wave and the inference of system statistics.  
By leveraging the  {\em Dynamic mode decomposition}, the pilot-wave dynamics has been shown to execute a period-doubling cascade typically observed in damped-driven systems ranging from detonation waves to mode-locked lasers.  DMD provides a low-rank approximation of the pilot wave dynamics into a set of spatial modes with associated temporal frequencies.  The regression framework of DMD provides a best-fit linear dynamics model over snapshots of spatio-temporal data.  The DMD characterization of the wave field yields a new perspective on the walking-droplet problem that forges valuable links with quantum mechanics. In particular, it naturally decomposes the wave field into modes of the form prevalent in standard quantum theory.

Our analysis has shown that as the vibrational acceleration is increased progressively, the pilot wave undergoes a series of Hopf bifurcations in which new modes at approximately harmonic frequencies emerge, culminating in a period-doubling cascade to spatio-temporal chaos. Such a period-doubling route to chaos is a canonical feature of damped-driven systems, which 
can be related to the logistic map in which the same canonical bifurcation structure is evident. The simplicity of the logistic map belies the rich and complicated behavior that it captures~\cite{may2004simple}.  Given the diversity of models capable of producing this same bifurcation structure~\cite{li2010geometrical,spaulding2002nonlinear,ding2009operating,koch2021multiscale}, it is highly suggestive that energy balance considerations alone can dictate the overall physics in such damped-driven systems.  Specifically, the gain and loss dynamics in the hydrodynamic pilot-wave system produce a mapping between the driving energy input and the damping losses that generate a period doubling cascade~\cite{rahman2022walking}.   

Our study has demonstrated how classical pilot-wave dynamics, like quantum mechanics, may yield predictions for the statistical behavior of particles on the basis of a wave theory. The map between the quantum and classical pilot-wave descriptions of particle motion in a one-dimensional well is presented in Table 1. In quantum mechanics, the relevant wave form is the wavefunction $\Psi$, while in our system it is the pilot wave. In quantum mechanics, the number of modes excited depends on the particle energy. In our pilot-wave system, the system memory plays an analogous role. Note that as either of these control parameters is increased progressively, a discrete set of new wave modes are introduced. In quantum mechanics, the waves correspond to the complex wavefunction $\Psi$; in our system, to the pilot wave. In quantum mechanics, the statistical behavior of the particles is prescribed by Born's Rule: the density of states is prescribed by the square of the wavefunction
$|\Psi|^2$. In our pilot-wave system, the density of states is inferred from the mean pilot wave via the operator inversion suggested by the theorem of 
Durey et al.~\cite{Durey2020_2}.

%[COMMENT - IT WOULD BE NICE TO CIRCLE BACK TO EQN(1), AND DISCUSS THE DIFFERENCE BETWEEN OUR NUMERICAL FORMALISM, AND THE ANALYTICAL FORMALISM OF QM.]

In summary, the DMD algorithm is a regression to exponential solutions of the form (\ref{eq:DMDapprox}). In quantum mechanics, the solution (\ref{eq:DMDapprox}) is typically constructed using analytical (or semi-analytical) techniques where appropriate boundary and matching conditions are imposed for a given potential.  Since Schr\"odinger's equation is linear, such an eigen-decomposition provides a set of linear basis modes which can be super-imposed to express any solution.  DMD is a regression directly to the solution form (\ref{eq:DMDapprox}), using variable projection to compute amplitudes $b_j$, modes $\phi_j$ and frequencies $\omega_j$ of the exponential solutions directly from data. In the quantum scenario, one must know the potential in order to construct the solutions.  In contrast, DMD can be used in a completely data-driven manner without knowledge of the underlying potential.  Thus DMD is advantageous for scenarios where an unknown, effective potential determines the underlying dynamics.  Moreover, it provides the best approximation of the form (\ref{eq:DMDapprox}) even when the underlying dynamics are nonlinear, such as is the case with the hydrodynamic analog system.              

\section*{Acknowledgements} PJB and JNK acknowledge support from the National Science Foundation AI Institute in Dynamic Systems (grant number 2112085). JWMB gratefully acknowledges support from the National Science Foundation through grant CMMI-2154151. AN acknowledges support from CNPq under  (PQ1D) 307078/2021-3  and
FAPERJ {\it Cientistas do Nosso Estado} project E-26/201.156/2021.

\section*{Appendix:  Calculating the particle PDF from the average wave field}
In this section, we detail the practical steps involved in computing the particle probability density function (PDF)  from the average wave field.
We use the full time window of the simulated system (\ref{Bern}--\ref{DropODE}) to calculate the average wave field and the true PDF.
Discretizing \eqref{Eq:durey} on a grid of equi-spaced points $\{x_1,\dots, x_{128}\}$ produces the system
\begin{align}
\overline{\bfeta}  = \bG \bmu,
\label{Eq:system}
\end{align}
where $\{\bmu\}_i = \mu(x_i)$, $\{\overline{\bfeta}\}_i = \overline{\eta}(x_i)$, and $\{\bG\}_{i,j} = \eta_G(x_i, x_j)$.
Our goal is now to find a $\bmu$ that solves \eqref{Eq:system} for a given Green's function matrix $\bG$ and an observed mean wave field $\overline{\bfeta}$.
The wave field is observed over a finite time horizon, so the resulting time average invariably contains noise. As such, the observed average wave differs from the true average wave so we write
\begin{align*}
    \overline{\bfeta} = \overline{\bfeta}^{\rm true} + \boldsymbol{\epsilon}
\end{align*}
where $\boldsymbol{\epsilon}$ is a vector of noise corresponding to the temporal truncation error.
Therefore, we consider a least-squares optimization problem to find the particle histogram: we seek the optimal vector $\bmu^\ast$
defined by
\begin{align}
\bmu^\star = \argmin_{\bmu} \| \bG \bmu - \overline{\bfeta} \|_2^2.
\label{Eq:opt}
\end{align}
Additionally, we are only interested in solutions that represent probability density functions.
In other words, $\bmu^\star$ must be non-negative and have unit mass, specifically,
\begin{align}
    \bmu^\star \geq \boldsymbol{0}, \qquad \qquad
    \|\Delta x \, \bmu^\star\|_1 = 1
    \label{Eq:constraints}
\end{align}
where $\Delta x = x_2-x_1$ is the equi-spaced quadrature weight.
The above constraints effectively regularize the problem \eqref{Eq:opt} and remove unphysical solutions.
%Finding a suitable $\bmu$ that solves \eqref{Eq:opt} subject to the constraints \eqref{Eq:constraints} presents several practical difficulties.
However, the matrix $\bG$ is ill-conditioned (with condition number $O(10^{10})$) so taking an inverse or repeatedly applying $\bG$ can be unreliable.
%[PB - PLEASE DEFINE KAPPA.]
Furthermore, $\bG$ has a large approximate nullspace so many candidate PDFs produce small residuals in \eqref{Eq:opt}.

We deployed several classical methods \cite{Kaipio2005} to solve the inverse problem in \eqref{Eq:opt}.
Ultimately, we found that the non-negative flexible conjugate-gradient least squares (NN-FCGLS) method of Gazzola and Wiaux \cite{Gazzola2017} provided the most robust results.
NN-FCGLS is a Krylov subspace method and is implemented in the IR Tools package \cite{Gazzola2019}.
Although NN-FCGLS does not enforce the unit mass constraint the computed solutions have mass very close to 1.

\begin{figure}[t]
\vspace{.5cm}
    \centering
\includegraphics[width = .95\linewidth]{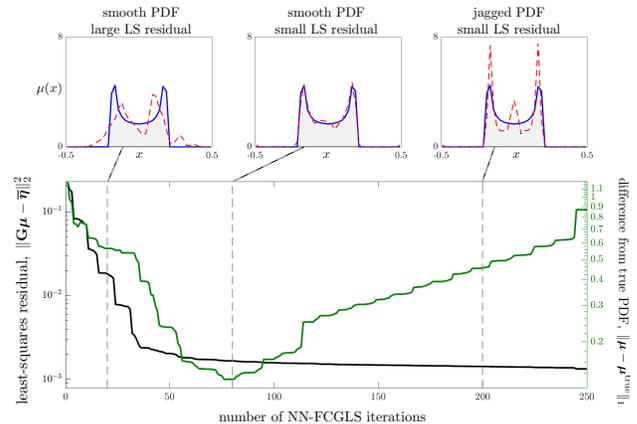}
    \caption{The effects of terminating the NN-FCGLS algorithm at different numbers of iterations.
    If the algorithm is terminated too early then the least-squares residual in \eqref{Eq:opt} is large; if the algorithm is terminated too late then the ensuing PDF is too jagged. 
    The plotted PDFs on the top row are computed with 20, 80, and 200 iterations respectively.
    }
    \label{fig:iterations}
\end{figure}

It is important to control the number of iterations executed by NN-FCGLS, as illustrated in figure \ref{fig:iterations}.
Therein, we plot the least-squares residual ($\|\bG \bmu - \overline{\bfeta} \|_2$)
and the error between the calculated and the true PDFs ($\| \bmu - \bmu^{\textrm{true}}\|_1$) when $\bmu$ is calculated after the given number of iterations.
In practice, the true PDF is unavailable so we include the PDF error here only for the sake of illustration.
We select the iteration number that produces a PDF satisfying two criteria: it must attain a small residual error in \eqref{Eq:opt}, and it must be sufficiently smooth.
Too few iterations can result in large residuals in \eqref{Eq:opt} and
too many iterations result in a jagged (non-smooth) solution.
Since the data is generated from a continuous dynamical system, we can exclude solutions that are not smooth.
Chapter 2 of \cite{Kaipio2005} suggests alternative criteria for selecting the number of iterations of conjugate gradient methods.

Figure \ref{fig:iterations} shows three candidate PDFs that we evaluate according to these two criteria.
The first PDF is produced after 20 NN-FCGLS iterations. It is smooth, but has a low least-squares residual so we discard it.
The third PDF is produced after 200 iterations. It has a small residual, but is jagged, so we discard it.
The second PDF is produced after 80 iterations. It has a small residual and is smooth. Accordingly, we select this PDF as our solution $\bmu^\star$. The green curve in figure \ref{fig:iterations} confirms that this is a good choice of PDF. Indeed, selecting anywhere between 70-100 NN-FCGLS iterations would have produced a satisfactory PDF.

Finally, an advantage of using Krylov methods is that they can be applied when one has access only to a linear operator that evaluates matrix-vector products $\bG \bv$, as opposed to $\bG$ itself. 
Accordingly, the NN-FCGLS method scales well and will be suitable for higher-dimensional problems, such as a particle in a circular corral \cite{Harris2013a,Durey2020_2}.

\section*{References}
\bibliographystyle{unsrt}
\bibliography{merged}

\end{document}